\documentclass[a4paper,11pt]{article}
\pdfoutput=1 

\usepackage{jcappub} 

\usepackage{longtable}
\usepackage{lscape}
\usepackage{subfigure}
\usepackage{multirow}
\usepackage{graphicx}
\usepackage{url}
\usepackage[latin1]{inputenc}

\newcommand{\lsim}{\mathrel{\mathop{\kern 0pt \rlap
  {\raise.2ex\hbox{$<$}}}
  \lower.9ex\hbox{\kern-.190em $\sim$}}}
\newcommand{\gsim}{\mathrel{\mathop{\kern 0pt \rlap
  {\raise.2ex\hbox{$>$}}}
  \lower.9ex\hbox{\kern-.190em $\sim$}}}

\newcommand{\alt}{\mathrel{\mathop{\kern 0pt \rlap
  {\raise.2ex\hbox{$<$}}}
  \lower.9ex\hbox{\kern-.190em $\sim$}}}
\newcommand{\agt}{\mathrel{\mathop{\kern 0pt \rlap
  {\raise.2ex\hbox{$>$}}}
  \lower.9ex\hbox{\kern-.190em $\sim$}}}

\newcommand{\gagamma}{g_{a\gamma}}
\newcommand{\ckcs}{counts~keV$^{-1}$~cm$^{-2}$~s$^{-1}$}

\title{Gaseous time projection chambers for rare event detection: Results from the T-REX project. II. Dark matter}

\author[a]{I.~G.~Irastorza,}
\author[a,1]{F.~Aznar,\note{Present address: Centro Universitario de la Defensa, Universidad de Zaragoza. Ctra. de Huesca s/n, 50090, Zaragoza, Spain}}
\author[a]{J.~Castel,}
\author[a]{S.~Cebri\'an,}
\author[a]{T.~Dafni,}
\author[a]{J.~Gal\'an}
\author[a]{J.~A.~Garcia,}
\author[a]{J.~G.~Garza,}
\author[a, 2]{H.~G\'omez,\note{Present address: APC, Univ Paris Diderot, CNRS/IN2P3, CEA/Irfu, Obs de Paris, Sorbonne Paris Cité, France}}
\author[a]{D.~C.~Herrera}
\author[a]{F.~J.~Iguaz,}
\author[a]{G.~Luzon,}
\author[a]{H.~Mirallas,}
\author[a]{E.~Ruiz,}
\author[a, 3]{L.~Segu\'i, \note{Present address: University of Oxford, Denys Wilkinson Building, Keble Road, Oxford, OX1 3RH, UK}}
\author[a, 4]{A.~Tom\'as,\note{Present address: The Blackett Laboratory, Imperial College London, UK}}

\affiliation[a]{Grupo de F\'{\i}sica Nuclear y Astropart\'{\i}culas, Departamento de Física Teórica, \\ Universidad de Zaragoza C/ P. Cerbuna 12 50009, Zaragoza, Spain}

\emailAdd{igor.irastorza@cern.ch}
\emailAdd{faznar@unizar.es}
\emailAdd{jfcastel@unizar.es}
\emailAdd{scebrian@unizar.es}
\emailAdd{tdafni@unizar.es}
\emailAdd{javier.galan.lacarra@cern.ch}
\emailAdd{jagarpas@unizar.es}
\emailAdd{jgraciag@unizar.es}
\emailAdd{hgomez@apc.univ-paris7.fr}
\emailAdd{dcherreramu@gmail.com}
\emailAdd{iguaz@unizar.es}
\emailAdd{luzon@unizar.es}
\emailAdd{mirallas@unizar.es}
\emailAdd{elisaruizcholiz@gmail.com}
\emailAdd{laurasgii@gmail.com}
\emailAdd{alfredo.tomas@gmail.com}

\abstract{As part of the T-REX project, a number of R\&D and prototyping activities have been carried out during the last years to explore the applicability of gaseous Time Projection Chambers (TPCs) with Micromesh Gas Structures (Micromegas) in rare event searches like double beta decay, axion research and low-mass WIMP searches. While in the companion paper we focus on double beta decay, in this paper we focus on the results regarding the search for dark matter candidates, both axions and WIMPs. Small (few cm wide) ultra-low background Micromegas detectors are used to image the axion-induced x-ray signal expected in axion helioscopes like the CERN Axion Solar Telescope (CAST) experiment. Background levels as low as 0.8 $\times 10^{-6}$ \ckcs have already been achieved in CAST while values down to $\sim10^{-7}$ \ckcs have been obtained in a test bench placed underground in the Laboratorio Subterráneo de Canfranc (LSC). Prospects to consolidate and further reduce these values down to $\sim10^{-8}$ \ckcs will be described. Such detectors, placed at the focal point of x-ray telescopes in the future International Axion Observatory (IAXO), would allow for 10$^5$ better signal-to-noise ratio than CAST, and search for solar axions with $\gagamma$ down to few 10$^{12}$~GeV$^{-1}$, well into unexplored axion parameter space. In addition, a scaled-up version of these TPCs, properly shielded and placed underground, can be competitive in the search for low-mass WIMPs. The TREX-DM prototype, with $\sim$0.300~kg of Ar at 10~bar, or alternatively $\sim$0.160~kg of Ne at 10~bar, and energy threshold well below 1~keV, has been built to test this concept. We will describe the main technical solutions developed, as well as the results from the commissioning phase on surface. The anticipated sensitivity of this technique might reach $\sim10^{-44}$~cm$^2$ for low mass ($<10$ GeV) WIMPs, well beyond current experimental limits in this mass range. }


\begin{document}
\maketitle
\flushbottom

\section{Introduction}

The development of gaseous Time Projection Chambers (TPCs) for the search of ``rare events'' has been the generic aim of the ERC-funded T-REX project since 2009. The motivation of this development has already been presented in the introduction of the companion paper~\cite{trexbbreview}. Most of the T-REX activity has been focused on Micromegas readouts~\cite{Giomataris:1995fq}, a type of micro-pattern gas detector (MPGD) that can be manufactured with extremely low level of intrinsic radioactivity, that have been introduced in section 2 of~\cite{trexbbreview}. This activity has opened promising prospects in the fields of double beta decay, axions and low-mass WIMPs searches. While in~\cite{trexbbreview} the results relevant to the search of the DBD of $^{136}$Xe have been described, we here focus on the results relevant to the search for two of the most interesting dark matter candidates: axions and WIMPs.

There is nowadays compelling evidence, from cosmology and astrophysics, that most of the matter of the Universe is in the form of non-baryonic cold dark matter (DM) \cite{Bertone:2004pz}. The particle-physics nature of this matter, however, remains a mystery. The Weakly Interacting Massive Particle (WIMP) is a good generic candidate to compose the DM. In addition, WIMPs appear naturally in well-motivated extensions of the Standard Model, e.g. the neutralino of supersymmetric (SUSY) extensions of the SM~\cite{Jungman:1995df}. Axions, on the other hand, are particles that appear in extensions of the Standard Model (SM) implementing the Peccei-Quinn (PQ) mechanism~\cite{Peccei:1977hh,Peccei:1977ur,Weinberg:1977ma,Wilczek:1977pj}, the most compelling solution to the long-standing strong-CP problem. They can be produced non-relativistically in the early Universe~\cite{Sikivie:2006ni,Wantz:2009it}, and thus they are also a favoured candidate to solve the DM problem. More generic axion-like particles (ALPs) appear in diverse extensions of the SM (e.g., string theory~\cite{Arvanitaki:2009fg,Cicoli:2012sz,Ringwald:2012cu}). ALPs could also be produced in the early Universe and be part of the DM~\cite{Arias:2012az}. Both axions and ALPs are repeatedly invoked to explain a number of poorly understood astrophysical observations (see e.g.~\cite{Irastorza:1567109} and references therein). It is important to stress that both axions (or the PQ mechanism) and WIMPs (or SUSY) are independently motivated by theory and they should not be regarded as alternative exclusive solutions of the DM problem. Axions and WIMPs (say, neutralinos) could both exist independently of being the dominant component of the DM. Moreover, the possibility of a mixed axion-WIMP DM is a scenario that has additional attractive features~\cite{Baer:2011uz,Bae:2013pxa}.

If our galactic DM halo is made of WIMPs, they could interact with nuclei and produce detectable nuclear recoils in the target material of underground terrestrial experiments. Due to the extremely low rate and low energy of those events, the experimental challenge in terms of background rate, threshold and target mass is formidable. During the last 30 years an ever growing experimental activity has been devoted to the development of detection techniques that have achieved increasingly larger target masses and lower levels of background, in the quest of reaching higher sensitivity to DM WIMPs. At the moment, the leading WIMP experiments are reaching sensitivities to the WIMP-nucleon cross section of $\sigma_N \sim10^{-45}$~cm$^2$ for WIMP masses $M_W \sim 50$~GeV, thanks to background levels of only a few counts per year for target masses at the $\sim$100~kg scale. Such impressive numbers are obtained thanks to the availability of discrimination techniques that allow distinguishing --with some efficiency-- electron recoils (produced e.g. by gammas) from the signal-like nuclear recoils.

Despite the enormous progress in WIMP experiments during the last 10--15 years, which has witnessed an improvement in sensitivity to $\sigma_N$ of more than 4 orders of magnitude, no convincing WIMP positive signal has been seen so far. This fact, together with the non-observation of signals of SUSY in the last run of the Large Hadron Collider (LHC) has triggered the revision of a number of working assumptions and the study of more generic phenomenological WIMP frameworks (other WIMPs interactions, different WIMP velocity distributions, etc.). As part of this attitude, interest in the search for low-mass WIMPs (i.e. below $\sim$ 10~GeV), a region of the WIMP parameter space poorly explored by mainstream experiments, is growing. The detection of low-mass WIMPs poses particular experimental challenges, because the energy deposits typically lie below the energy threshold of the discrimination techniques of mainstream experiments. As will be argued in this paper, gaseous TPCs might play a relevant role in this particular quest. The results of T-REX in this respect are presented from section \ref{sec:trexdm} onwards.

In addition, in part due to the lack of positive WIMP and SUSY signals, the search for axions is receiving increasing attention lately. However, axions are being searched for since their proposal more than 30 year ago. The detection of axions is conceptually very different than WIMP detection~\cite{ANDP:ANDP201300727,2015ARNPSAxions}. It relies on the fact that axions are predicted to convert into photons (and viceversa) in the presence of electromagnetic fields~\cite{Turner:1989vc}, a process that is sometimes referred to as Primakoff effect, in analogy with the pion-photon conversion. If DM is made of axions, a tiny microwave signal is expected to appear in magnetic resonant cavities~\cite{Sikivie:1983ip}, a detection concept known as ``axion haloscopes'', and technologically very different that typical particle detectors. However, if axions exist (and independently on whether they are part of DM or not) they must also be copiously produced at the Sun's core. Axion emission by the solar core is a robust prediction involving well known solar physics and the Primakoff conversion of plasma photons into axions. Solar axions have $\sim$keV energies and in strong laboratory magnetic fields can convert back into detectable x-ray photons. This detection concept is called ``axion helioscope''~\cite{Sikivie:1983ip}. The basic layout of an axion helioscope thus requires a powerful magnet coupled to one or more x-ray detectors. When the magnet is aligned with the Sun, an excess of x-rays at the exit of the magnet is expected, over the background measured at non-alignment periods. Therefore, ultra-low background x-ray detectors are a technological pillar of axion helioscopes~\cite{Irastorza:2011gs}.

The currently most powerful axion helioscope is the CERN Axion Solar Telescope (CAST), now in operation for more than a decade at CERN. The use of small Micromegas TPCs as x-ray detectors in the CAST experiment dates back to 2003 and it is indeed a pioneer use of Micromegas in a rare event search~\cite{Abbon:2007ug}. This early work did motivate the later T-REX development whose results are here reviewed. Since then, CAST has been a test bench for these developments. Currently, CAST Micromegas detectors have demonstrated the best signal-to-noise figures in the history of CAST, and are the baseline detection technology for the future International Axion Observatory (IAXO)~\cite{Irastorza:2011gs,Armengaud:2014gea}, the next generation axion helioscope under proposal.

The present paper is organized as follows. In section \ref{sec:MMaxion} we review the role of Micromegas TPCs in axion helioscopes, and we present the basic layout of x-ray detectors developed for CAST, along with the low-background techniques used in their conception. In section \ref{sec:lowb} we review the status of the R\&D so far including levels of background and energy threshold obtained by the latest prototypes, and in particular the experience in the so-called IAXO pathfinder system in operation at CAST. In section \ref{sec:iaxo} we discuss the physics prospect associated with the future IAXO experiment. In section \ref{sec:wimps} the motivation of gaseous TPCs in the search for low mass WIMPs is discussed. In section \ref{sec:trexdm} we describe the TREX-DM prototype built to test this concept, including the vessel, readout, and other main elements of the setup, as well as the latest results of the ongoing commissioning phase. In section \ref{sec:prospects} we briefly discuss the anticipated sensitivity prospects for low mass WIMPs using this detection concept. Finally, in section \ref{sec:conclusion} our conclusions and generic prospects are summarized.

\section{Micromegas chambers in the search for solar axions}
\label{sec:MMaxion}

\begin{figure}[t]
  \centering
\includegraphics[height=8cm]{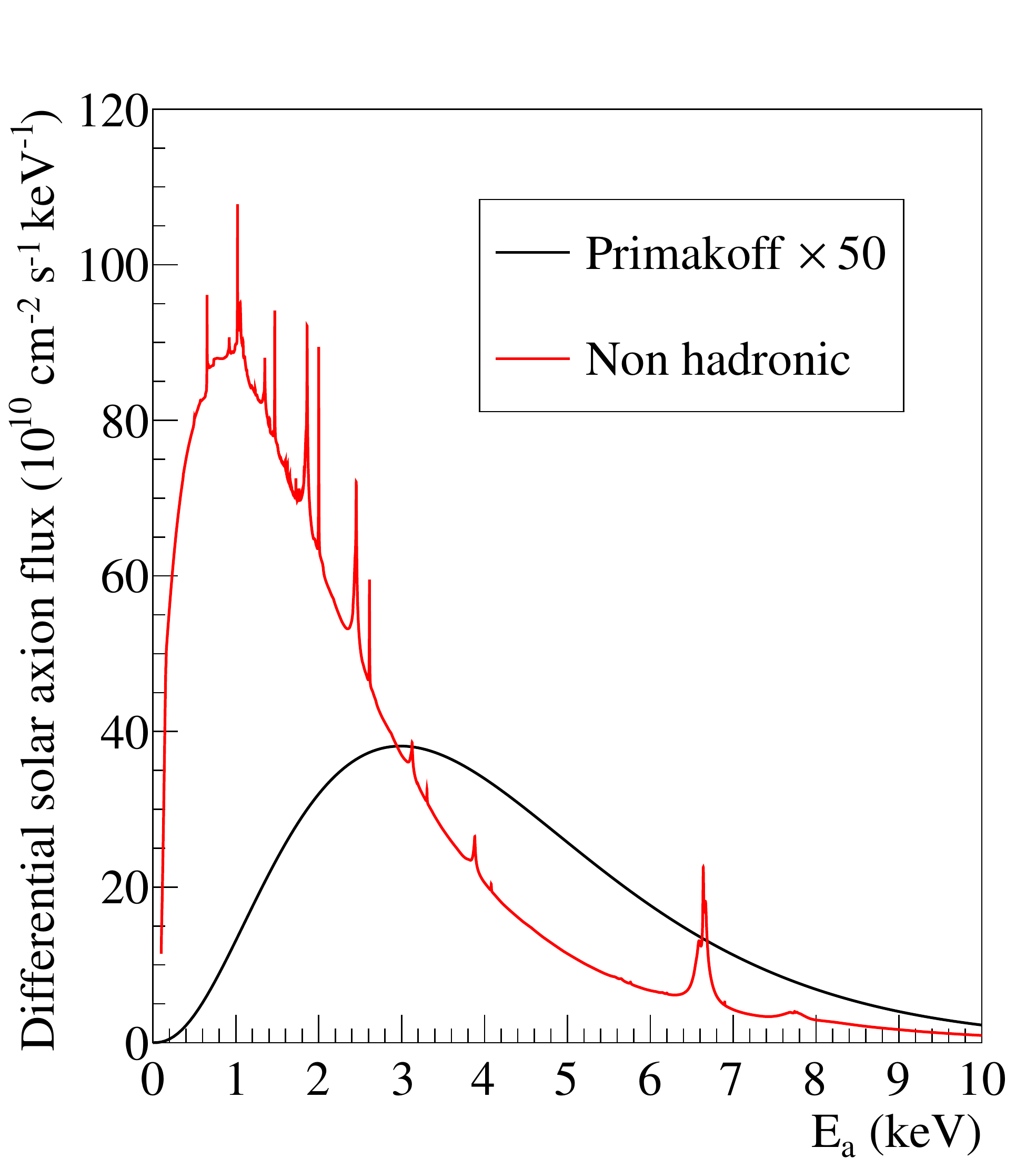}
 \caption{ Solar axion flux due to the standard Primakoff conversion (in black) for $\gagamma=10^{-12}$ GeV$^{-1}$, as well as from reactions involving electrons (in red) for axion-electron coupling $g_{ae}=~10^{-13}$~\protect\cite{Redondo:2013wwa}. The Primakoff spectrum has been scaled up by a factor 50 to make the two contributions comparable in the plot. }
\label{fig:axion_flux}
\end{figure}

Axion helioscopes make use of a large and powerful magnet aligned to the Sun. In its magnetic field solar axions are converted into x-rays. The energy distribution of these photons follows that of the solar axions, which typically lies at the 1$-$10~keV range \cite{Andriamonje:2007ew}, with a peak at $\sim$4~keV. In some axion models (if the axion couples with electrons at tree level) other solar axion production mechanisms are available \cite{Redondo:2013wwa}, and the spectrum is peaked at lower energies (see Fig.~\ref{fig:axion_flux}), and sub-keV x-ray sensitivity is required. The most powerful axion helioscope so far is the CERN Axion Solar Telescope (CAST), now in operation for more than a decade at CERN. It uses a Large Hadron Collider (LHC) dipole prototype magnet with a magnetic field of up to 9~T over a length of 9.3~m and an aperture of $2\times15~$cm$^2$~\cite{Zioutas:1998cc}, that is able to follow the Sun for $\sim$3 hours per day using an elevation and azimuth drive. During its 10-year experimental program~\cite{Zioutas:2004hi,Andriamonje:2007ew,Arik:2008mq,Aune:2011rx,Arik:2013nya,Arik:2015cjv}, CAST has seen no signal of solar axions, and a number of upper limits on the photon-axion coupling $\gagamma$ at the level of $10^{-10}$GeV$^{-1}$ for axion masses up to $\sim$1~eV have been derived. It is the most stringent experimental bound in most of this axion mass range.
More recently it has been proven~\cite{Irastorza:2011gs} that the helioscope technique can be substantially scaled up in size by building a large aperture superconducting magnet, and by extensive use of x-ray focusing optics and low background x-ray detection techniques. This has been the basis for the proposal of a next generation axion helioscope, the International Axion Observatory~\cite{Irastorza:1567109,Armengaud:2014gea}. Future prospects for solar axion searches in the context of IAXO will be discussed in section \ref{sec:iaxo}.

\begin{figure}[t]
\begin{center}
\includegraphics[height=6cm]{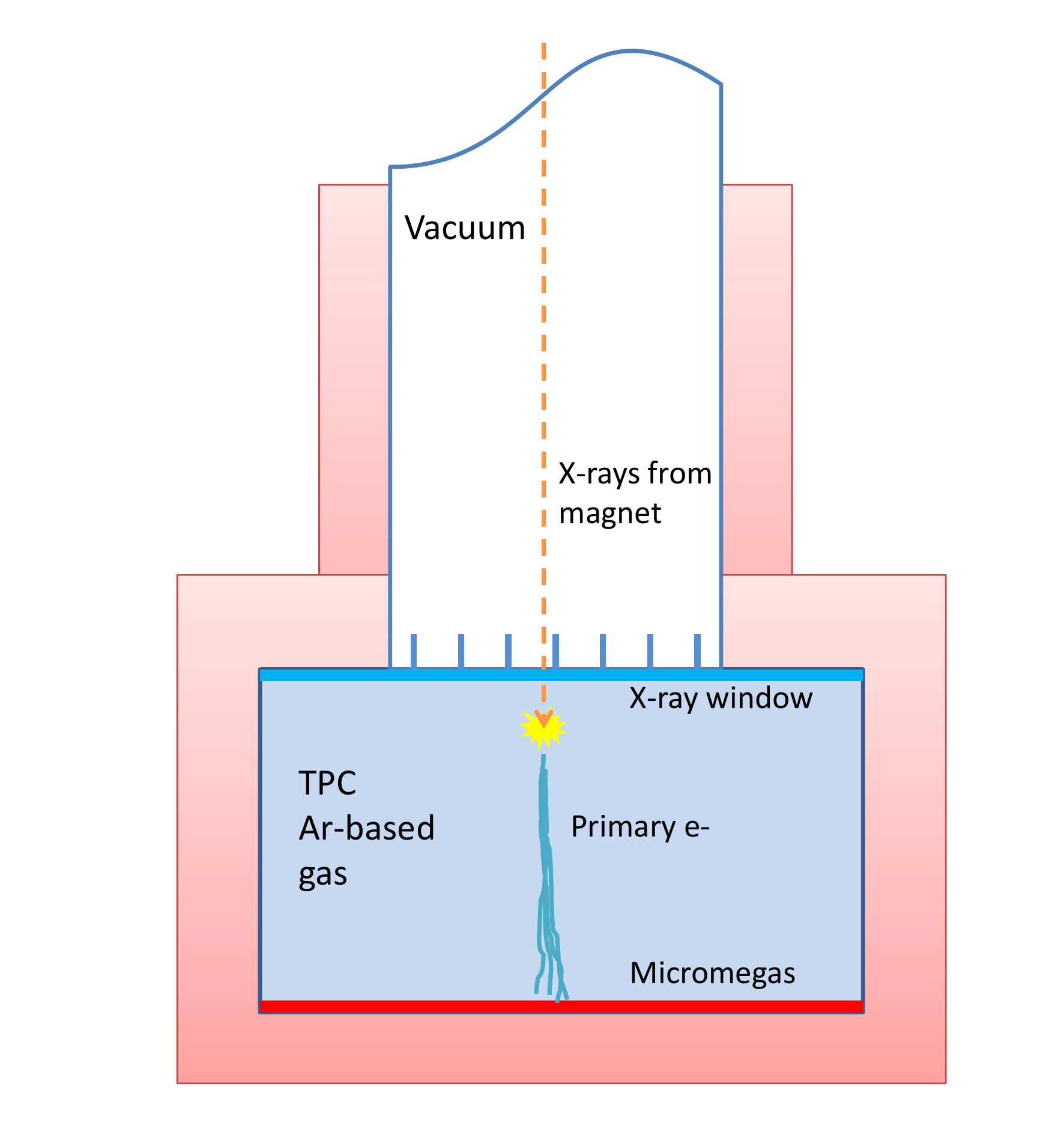}
\includegraphics[trim=1.5cm 0cm 1.5cm 0cm, clip=true,height=6cm]{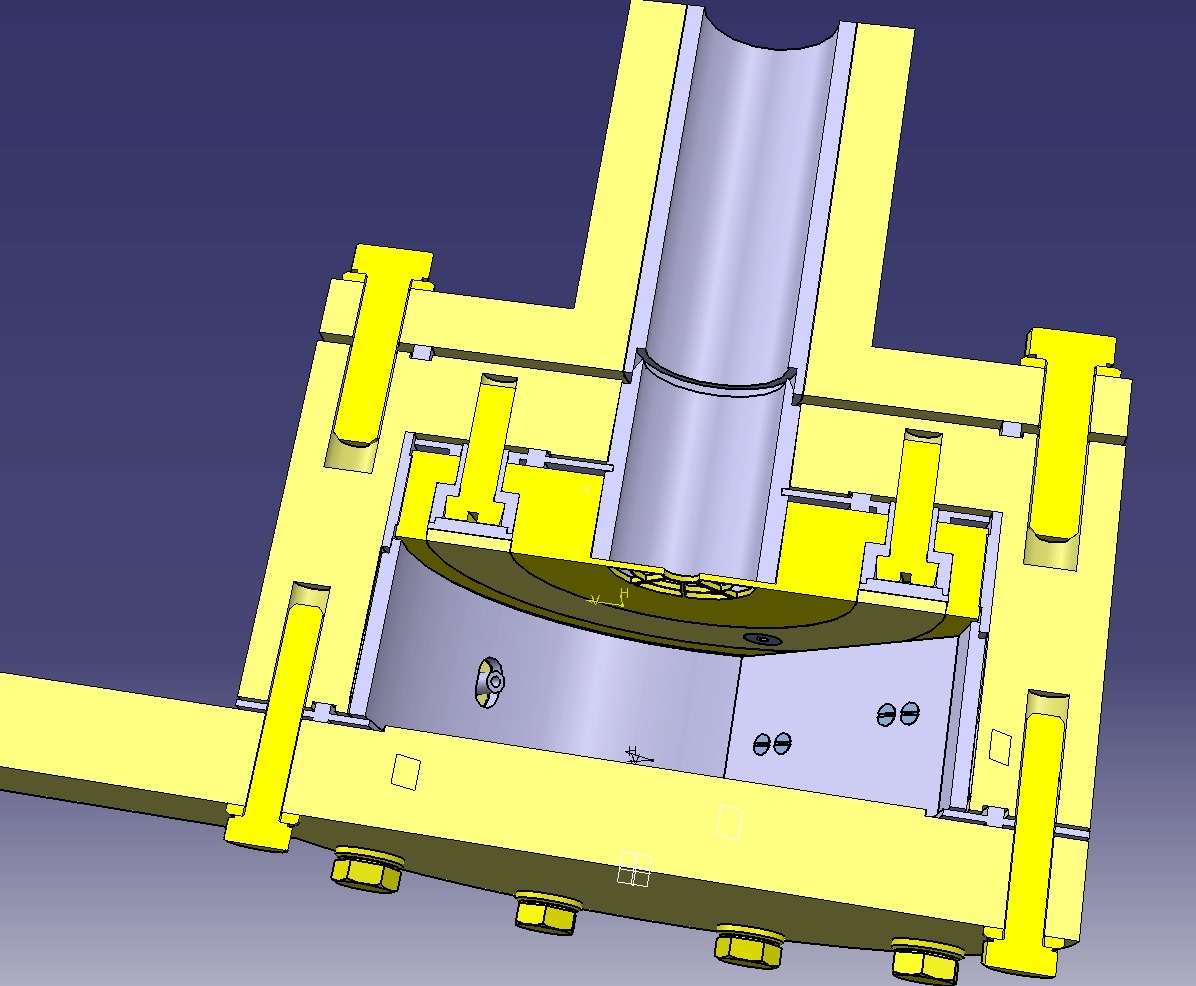}
\caption{\label{fig:det_concept} Left: Sketch of the detection principle of Micromegas detectors in IAXO. Right: Design of the IAXO detector prototype.}
\end{center}
\end{figure}

One of the experimental challenges of axion helioscopes is thus the detection of as low an x-ray flux as possible in the 1$-$10~keV energy range. For this, the use of low background x-ray detection techniques is needed, potentially coupled to x-ray optics to focus the parallel beam of photons into a small spot, in order to further increase the signal-to-noise ratio. The concept of detecting low energy x-rays with Micromegas detectors has been extensively developed and used in CAST since an early stage of the experiment~\cite{Abbon:2007ug,Aune:2013pna,Aune:2013nza}. Despite the many changes and improvements in the CAST Micromegas detectors over the years, the basic concept has remained unchanged and is illustrated by the sketches in Figure~\ref{fig:det_concept}. The detector is a small TPC with a Micromegas readout at the anode, the cathode of which faces the magnet bore from where signal x-rays enter the detector. The conversion volume of the chamber is fixed to efficiently stop the signal photons, while minimizing background, and typically has 3~cm height and is filled with 1.4~bar Argon in addition to a small quantity of quencher (e.g. 2\% isobutane).

The x-rays coming from the magnet enter the conversion volume via a gas-tight window made of 4~$\mu$m aluminized mylar foil. This foil is also the cathode of the TPC, and it is supported by a metallic strong-back, in order to withstand the pressure difference with respect to the magnet's vacuum system.

The ionization produced by interactions in the conversion volume drifts and projects onto the Micromegas readout plane, where signal amplification takes place. The basics of Micromegas operation as well as their radiopurity have already been described in the companion paper~\cite{trexbbreview}. The Micromegas active area is a square of about 60$\times$60~mm$^2$ that comfortably covers the 14.55~cm$^2$ projection of the magnet's circular bore, which fixes the analysis fiducial area (in the case of the optic-less detectors). The readout plane is finely pixelized with a typical pitch of $\sim$0.5~mm. The pixels are alternatively interconnected in $x$ and $y$ directions through metallized holes, routed to the connector prints at the edge of the detectors support and then driven through flat cables to the front-end electronics. Typically, half of the pixels are connected at the top anode layer, and the other half are connected in an underlying copper layer. The micromesh holes are arranged in groups that are aligned to the pixels underneath. The strips are connected to the detector ground via a 10~M$\Omega$ resistance that can be removed in case of mesh-strip short-circuit. The signals are read using state-of-the-art TPC data acquisition (DAQ) electronics based on the AFTER chip~\cite{Baron:2008zza,Baron:2010zz}. This provides relative time-of-flight information of the ionization charge arriving to the readout plane and hence 3D information of the primary ionization cloud. We refer to \cite{Aune:2013pna} for details on the technical implementation of these detectors in CAST.


The reasons why these detectors have achieved the best background levels in CAST are due to the following:

\begin{itemize}
  \item Micromegas readouts can be manufactured with very high standards of radiopurity, following the \textit{microbulk} technique. Microbulk readouts are made out of just kapton and copper, two materials of well known radiopurity (see section 3 of~\cite{trexbbreview}). In addition, microbulks offer the possibility to extract the primary signals out of the detector to the point where the front-end electronics is (that can be relatively far and shielded from the detector), in a relatively straightforward way which does not involve soldering, connectors or any other component, which are potential sources of radioactivity. All the relevant components (chamber's body, x-ray window, screws, gas gaskets, connectors, etc.) have gone through screening campaigns and are built from radiopure materials.
  \item The detailed imaging of the ionization signature in gas provides invaluable information on the nature of the interaction and therefore a way to identify signal-like events. Indeed, the information obtained by the patterned anode, and the digitized temporal wave-forms of the pulses, is the basis to develop advanced algorithms to discriminate signal x-ray events from any other type of events.
  \item The detectors can be surrounded by appropriate passive and active shielding to reduce external background sources. Although shielding concepts from underground experimentation can be borrowed, care must be paid to the specifics of our case, e.g., the space and weight constraints of the magnet moving platform, the operation at surface (presence of cosmic rays), the geometry imposed by the magnet (the shielding will always have an opening from which the signal x-rays reach the detector) and the intrinsic sensitivity and rejection capability of the Micromegas detectors.
\end{itemize}

These points have been the basis for the development carried out within T-REX, and of the results that are reviewed in the following sections.

\section{Low background x-ray detection with Micromegas}
\label{sec:lowb}

\begin{figure}[!t]
\centering
\includegraphics[width=0.8\textwidth]{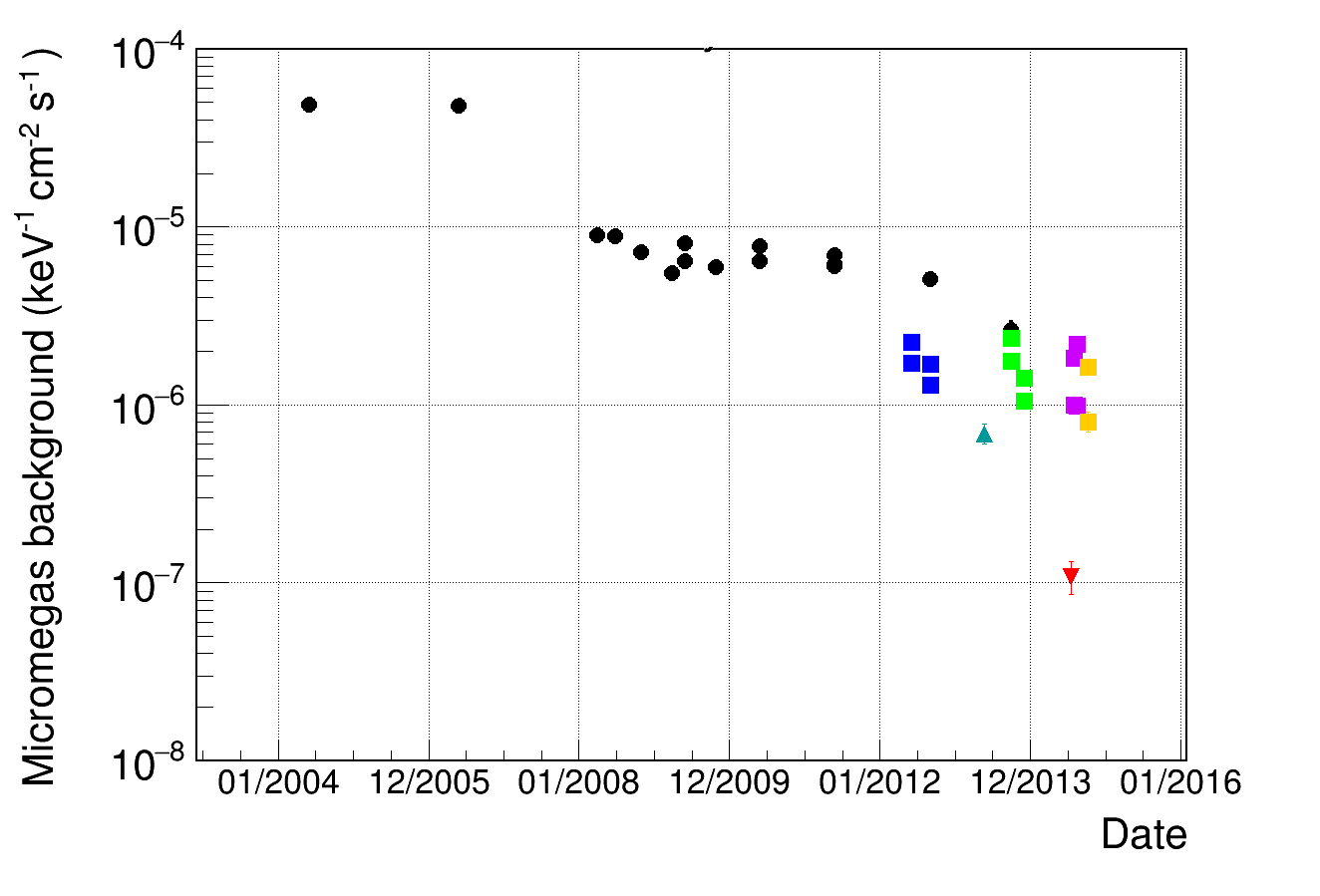}
\caption{Evolution of the background level in the Micromegas of CAST since the first detectors were installed in 2003. Black circles represent the values obtained in \emph{in-situ} measurements at CAST. Blue, green and purple squares represent the levels of the sunset detectors during the 2012, 2013 and 2014 data-taking campaigns, each pair of points representing the level before and after the application of the veto cut. Yellow squares are the levels obtained after the sunrise upgrade in 2014 before and after the application of the veto cut. The dark cyan triangle is the level obtained in a test bench at the Zaragoza laboratory, and the red triangle is the level obtained in a test setup operated at the Underground Canfranc Laboratory (LSC).}
\label{fig:historic}
\end{figure}

Since the beginning of the CAST experiment there has been at least one Micromegas chamber among the x-ray detectors of the experiment. Since 2008, three of its four detectors are microbulk Micromegas-based TPCs~\footnote{The CAST magnet is twin aperture, so two side-by-side detectors can be accommodated at each end of the magnet}. CAST has therefore served as test ground to apply the low background techniques developed for these detectors. Over the years, the CAST detection systems have been progressively improved, and the corresponding background levels reduced. Figure~\ref{fig:historic} shows the evolution of the background levels of the CAST detectors (during the experimental data taking campaigns) since the beginning of operations in 2003. This evolution responds to progressive and iterative understanding of background sources (supported by simulations and experimental data), followed by the improvement of detector components, shielding or offline discrimination strategies. This work has been the object of a series of technical papers~\cite{Abbon:2007ug,Aune:2013pna,Aune:2013nza,Aznar:2015iia} over the last decade or so. In the following we focus on the last series of detectors currently in operation in CAST, that exemplify the present state-of-the-art in background reduction at the few keV energies. As will be seen, this technique enjoys the best background levels ever achieved in detectors exposed to external x-ray radiation. In the next section we will briefly describe the status of the (partial) understanding of the origin of the current background level, that sets the path towards subsequent improvement, in view of application in the future IAXO experiment.

\begin{figure}[!t]
\centering
\includegraphics[width=0.65\textwidth]{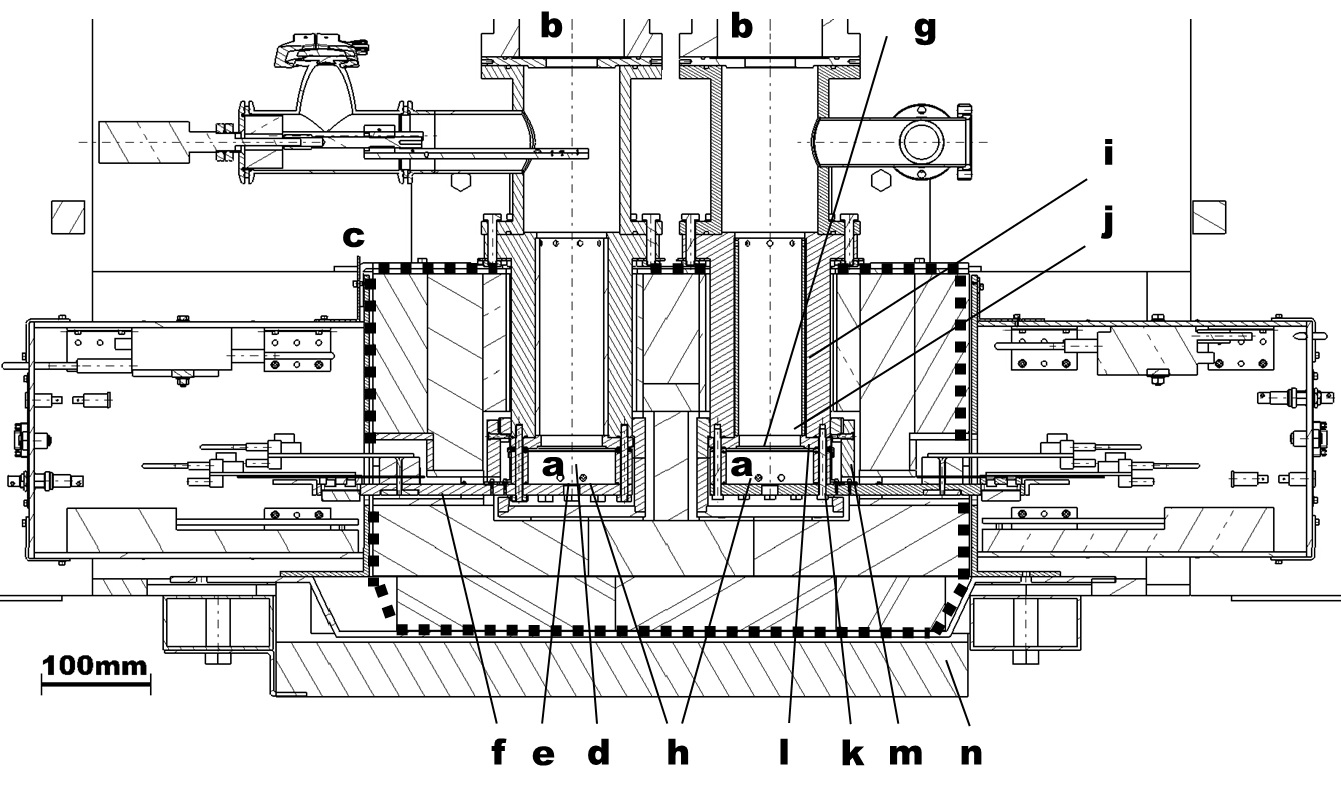}
\includegraphics[width=0.65\textwidth]{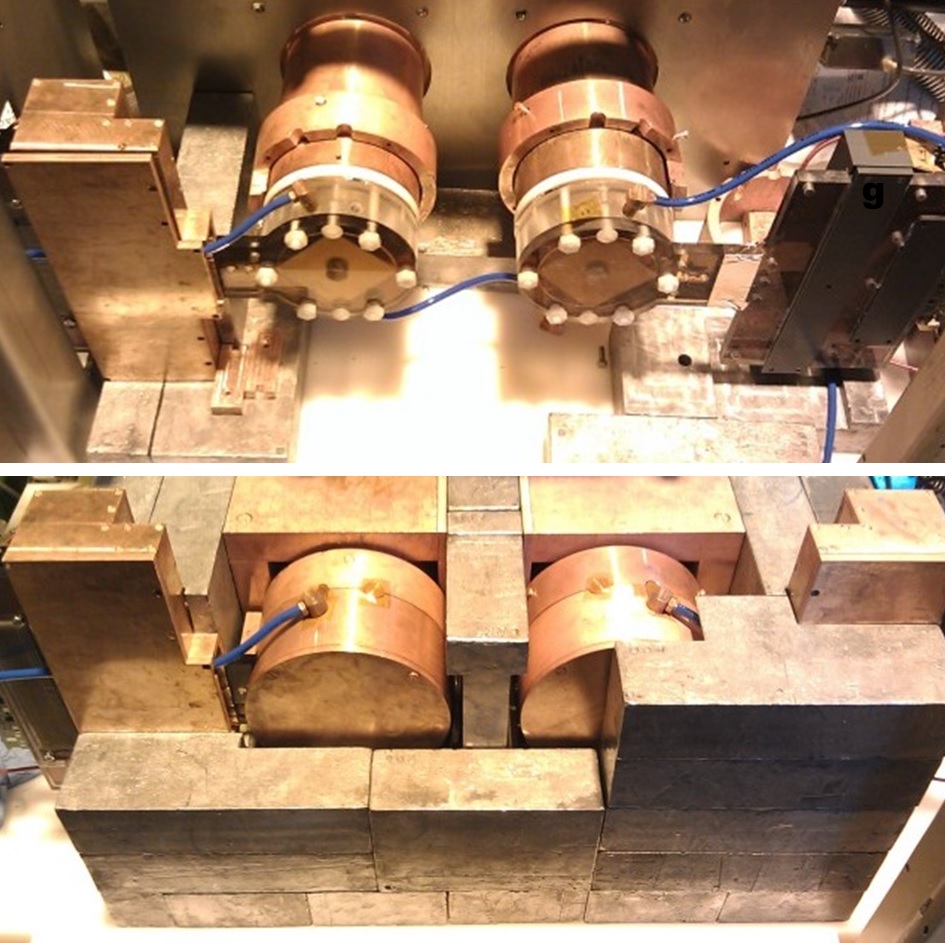}
\caption{Top: design (top view) of the CAST sunset system composed of two side-by-side detectors facing the magnet. The parts composing the system are labelled as: detector (a), magnet bore (b), lead shielding (c), plexiglass chamber (d), microbulk readout (e), \textit{raquette} (f), mylar window (g), gas ports (h), copper pipes (i), PTFE layer (j), nylon threaded bars (k), gasket (l), copper shielding (m), plastic scintillators	 (n). Middle and bottom: two pictures of the system without and with part of the shielding respectively.}
\label{fig:ssmmline}
\end{figure}

\begin{figure}[!t]
\centering
\includegraphics[width=0.45\textwidth]{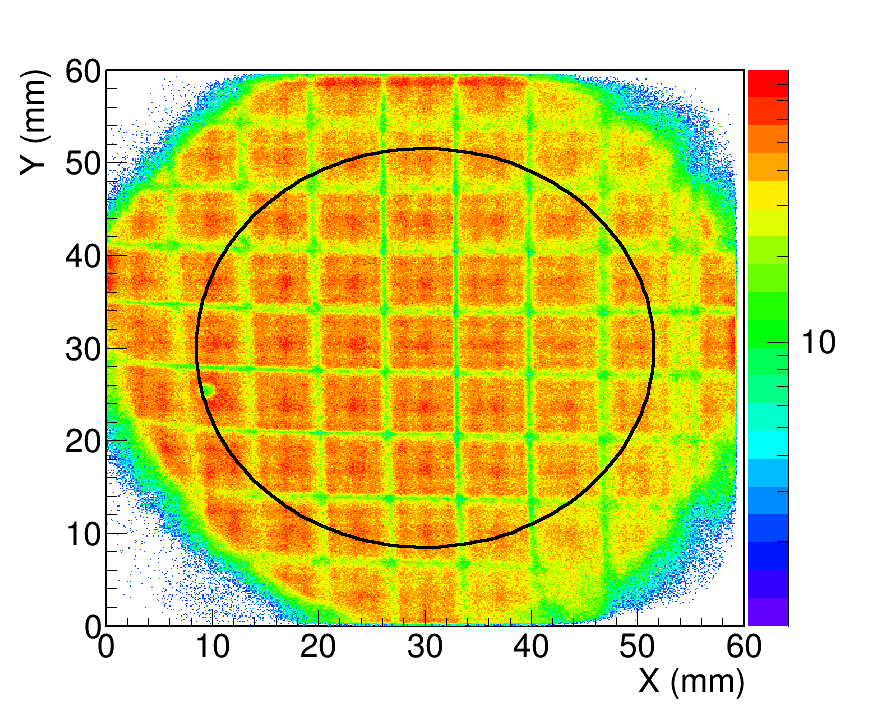}
\includegraphics[width=0.4\textwidth]{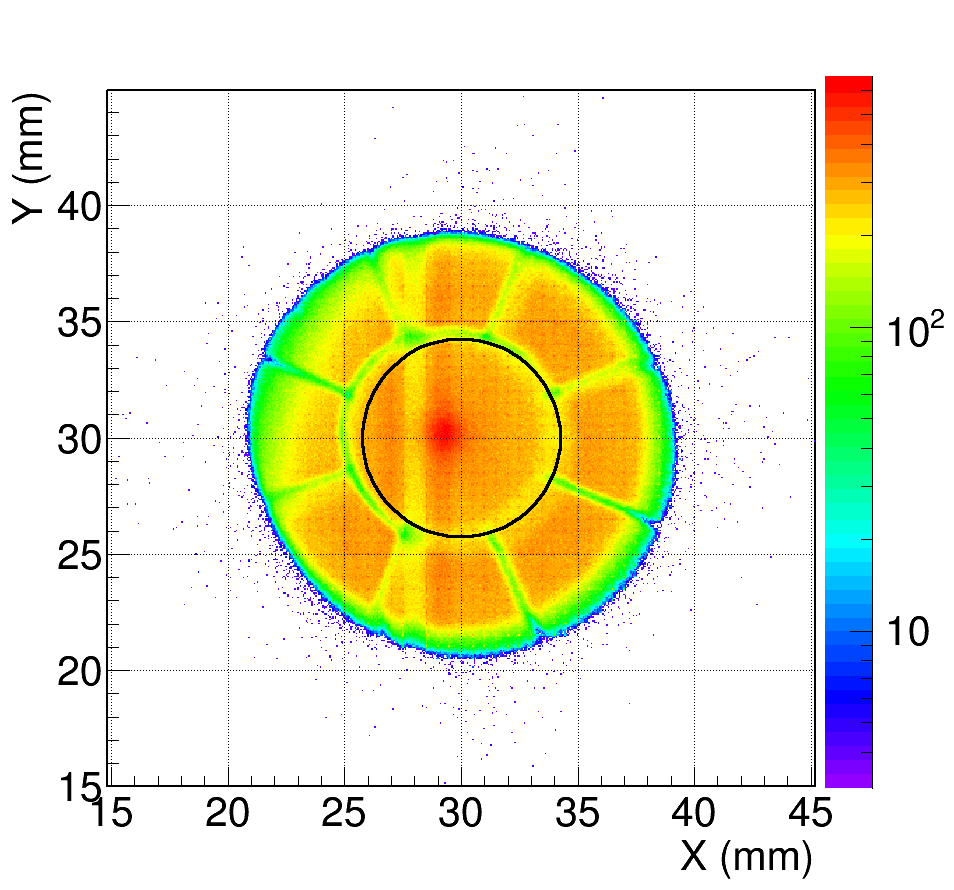}
\caption{Intensity maps produced by the $^{55}$Fe source on sunset (left) and sunrise (right) Micromegas detectors during the 2014 season. The black contours represent the cold-bore region and the inner circumference of the spider-web strong-back, respectively.}
\label{fig:2dmaps}
\end{figure}

Two of the referred microbulk CAST detectors are installed in the \textit{sunset} side  of the CAST magnet (i.e. sensitive to axions during the evening shifts of the experiment), while the third one is at the \textit{sunrise} end. The current configuration of the sunset system was defined in 2012, and since then has been in stable operation in the experiment. As shown in Figure~\ref{fig:ssmmline}, it is composed of two identical detectors (\textbf{a}), each of them facing one of the parallel magnet bores (\textbf{b}), and sharing a common shielding (\textbf{c}). The detector body is based on a plexiglass chamber (\textbf{d}), screwed on a plexiglass support (\textit{raquette}) where the microbulk readout is glued (\textbf{e}). The \textit{raquette} (\textbf{f}) extends well beyond the detector area, bringing the signal strips outside the shielding, avoiding the presence of connectors or soldering, close to the sensitive volume. The chamber holds a metallic squared grid (strongback) to which a thin 4~$\mu$m aluminized mylar x-ray window is glued (\textbf{g}). A 2D hitmap recorded by the detector during x-ray calibrations is shown in Figure~\ref{fig:2dmaps} (left), where the imprint of the cathode window strongback is clearly seen. The strongback and the window are the cathode of the TPC, biased to a voltage that defines the drift field. The x-ray window allows the passage of the x-rays and withstands the pressure difference between the chamber and the vacuum line. The chamber has two gas ports (\textbf{h}) that allow to operate the detector in gas circulation mode, normally at 1.4~bar of Ar+2\% iC$_{4}$H$_{10}$.  The overpressure guarantees a good quantum efficiency for x-rays in the energy region of interest (RoI), while the leak-tightness of the detection chamber is achieved by viton o-rings.

One of the important improvements over previous implementations is the removal of stainless steel fluorescences (Fe-K$_{\alpha}$ at 6.4~keV, Mn-K$_{\alpha}$ at 5.9~keV or Cr-K$_{\alpha}$ at 5.4~keV), which fall in the CAST energy RoI. The connection of the detectors to the magnet is done with 10~mm thick, 200~mm long, high purity copper pipes (\textbf{i}), coated in the inside with a polytetrafluoroethylene (PTFE) layer (\textbf{j}) to absorb the 8~keV copper fluorescence. The detectors are coupled to the copper tube via Nylon 66 threaded bars (\textbf{k}). A PTFE gasket (\textbf{l}) is used to preserve the vacuum tightness and to electrically isolate the vacuum tube and the cathode. A first shielding layer of 10~mm thick copper (\textbf{m}) surrounds the plexiglass body. The lead shielding extends approximately 100~mm all around the detectors, covering also the copper pipes. The only opening of the shielding is for the connection to the magnet and where the \textit{raquette} extracts the signals to the front electronics. Covering the top and back sides of the shielding, two plastic scintillators (\textbf{n}) act as muon vetos.

On the other hand, the new sunrise system was installed and commissioned during the 2013 and 2014 campaigns, and in some aspects it builds up and improves on the sunset detectors. Contrary to the latter, which are placed directly facing the magnet, the sunrise detector is placed at the focal point of an x-ray telescope, that focuses the signal photons onto a small spot area. The overall system has been developed as a technological pathfinder for IAXO, combining two of the techniques (optics and detector) proposed in the conceptual design of the project~\cite{Armengaud:2014gea}. The x-ray telescope is a cone-approximation Wolter I x-ray optic of $\sim$5~cm diameter (enough to cover the aperture of one CAST magnet bore) and 1.3~m focal-length. It is composed of thermally-formed glass substrates deposited with multilayer coatings. This technology for x-ray optics is the one used in the hard x-ray mission NuSTAR~\cite{Harrison:2013md} and was identified in \cite{Irastorza:2011gs} to have the potential to cost-effectively cover the areas needed for IAXO with sufficient performance. Although CAST already used x-ray focusing optics for one of its four detector lines (one of the spare optics from the ABRIXAS x-ray mission), this is the first time an x-ray optic is designed and built for an axion application~\cite{Aznar:2015iia}. It is also the first time a Micromegas detector is operated with x-ray optics. The full system was commissioned in September 2014 in CAST, and is since then in operation, having achieved the best signal-to-background ratio obtained so far in CAST.

\begin{figure}[!t]
\centering
\includegraphics[width=0.75\textwidth]{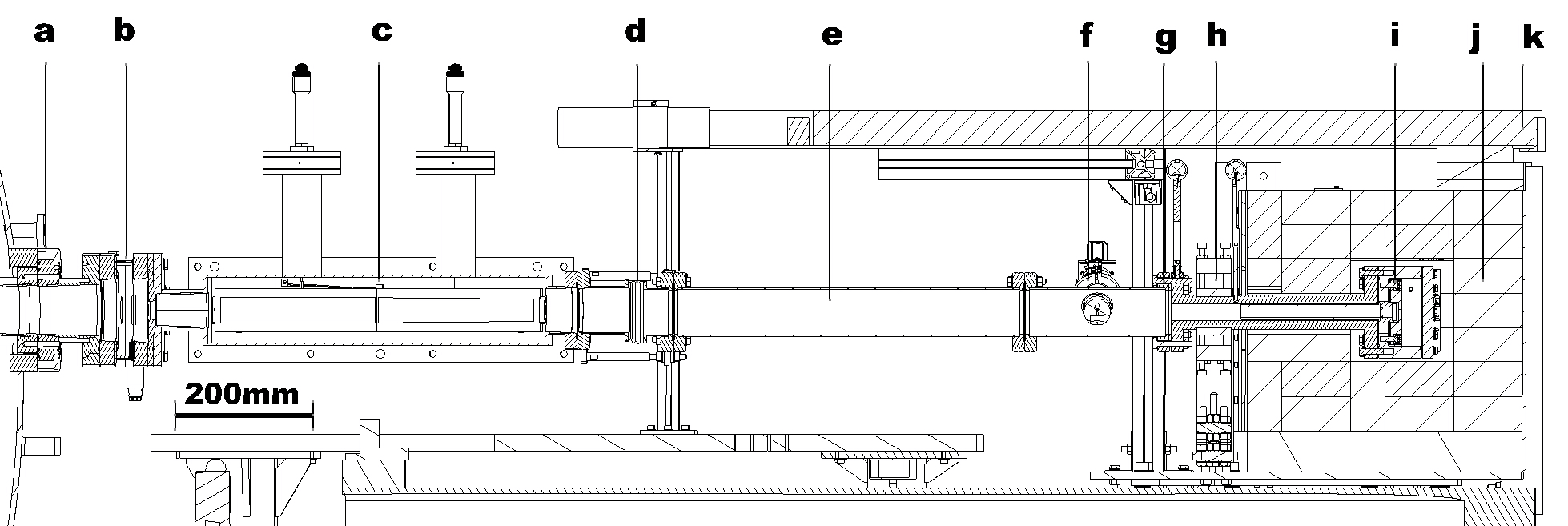}
\includegraphics[width=0.7\textwidth]{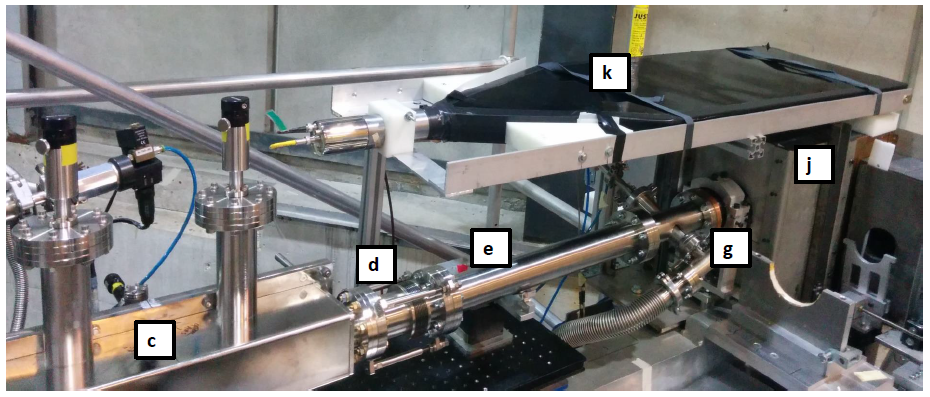}
\caption{Design (side view) and photo of the new CAST sunrise detection line composed of a low background Micromegas detector placed at the focal point of an x-ray optic. Signal photons come from the left of the figure (end of magnet). The different parts of the line are labelled in the sketch: gate valve (a), differential window (b), Wolter I x-ray optic (c), bellow (d), stainless steel interface tube (e), calibration system (f), precision stage (g), copper interface tube (h), Micromegas detector (i), lead shielding (j) and muon veto (k).}
\label{fig:srmmline}
\end{figure}

The sunrise Micromegas detector features some technical improvements over the ones in the sunset system. The chamber, Faraday cage and inner copper shielding are completely integrated in an \emph{all-in-one} piece, made of 18~mm thick radiopure copper. The \textit{raquette} is now also made in radiopure copper. As before, it serves as a support where the readout is glued and to extract the signals out of the shielding. A redesign of the routing on the microbulk foil, reducing all connections going out of the shielding into strips on the kapton foil (including all HV connections), allows to effectively close the opening in the shielding, while keeping all connectors to the front end electronics and power supplies outside. In addition, a new field shaper has been designed in order to improve the homogeneity of the drift field in the conversion volume. It is printed on a flexible polyimide multilayer. All the materials of the mechanical elements of the detector have been reduced to copper and PTFE. For example, all the gaskets are now made of PTFE instead of the previous elastomeric ones. Figure~\ref{fig:srmmline} shows the overall sunrise detection line, built to keep the detector (\textbf{i} in the sketch of Figure~\ref{fig:srmmline}) in the focal point of the optic (\textbf{c}), as well as to align both elements with the axis of the magnet bore. As in the case of the sunset system, the detector is surrounded by a lead shielding (\textbf{j}) and covered at the top by a plastic scintillator(\textbf{k}) as muon veto. 

Due to the telescope focusing, signal photons enter the detector over an area of only a few mm$^2$, much smaller than the magnet bore. This allows to reduce the diameter of the pipe connecting the detector with the optic (at least the part closest to the detector), and therefore to reduce the non-shielded solid angle of the setup. The x-ray window on the detector is accordingly smaller. Its strongback pattern has been modified to a spider-web design with a central hole of 8.5~mm of diameter, large enough to contain the expected axion signal image. The x-rays focused by the optic go through the 4~$\mu$m aluminized polypropylene window, avoiding the grid structure, whose opacity was responsible of a $\sim$10\% of efficiency loss in previous setups. A 2D hitmap recorded by the detector during x-ray calibrations is shown on the right of Figure~\ref{fig:2dmaps}. It clearly shows the imprint of the spider-web strongback. The inner circle of the cathode strongback (marked with the overlaid black circle) should comfortably include the focal spot area where the axion signal is expected. The exact knowledge of the position of the spot is crucial for the experiment. The spot position has been measured with both a dedicated in-situ laser alignment, as well as periodically monitored with an x-ray generator from the other side of the magnet. A detailed description of the commissioning, alignment and spot-calibrations can be found in reference~\cite{Aznar:2015iia}.

The sources of background for the Micromegas x-ray detectors operating at surface are, generally: $\gamma$-rays from the environmental radiation, cosmic rays, presence of radon around the detector, intrinsic radioactivity of the detector components or shielding, neutrons and cosmogenic activation of the detector materials. All these radiations can produce secondary or fluorescence emissions that can reach the detector active volume. The generic strategies to reduce these backgrounds include: the use of high-Z material (lead, copper) passive shielding to block the passage of $\gamma$-rays; active shielding to tag muons, such as plastic scintillators; the continuous flux of vapourized LN$_2$ into the detector environment to avoid or reduce the presence of air-borne radon; a low-Z material to moderate neutrons plus a layer of a neutron absorber; and the use of radiopure components.

\begin{figure}[!t]
\centering
\includegraphics[height= 6cm]{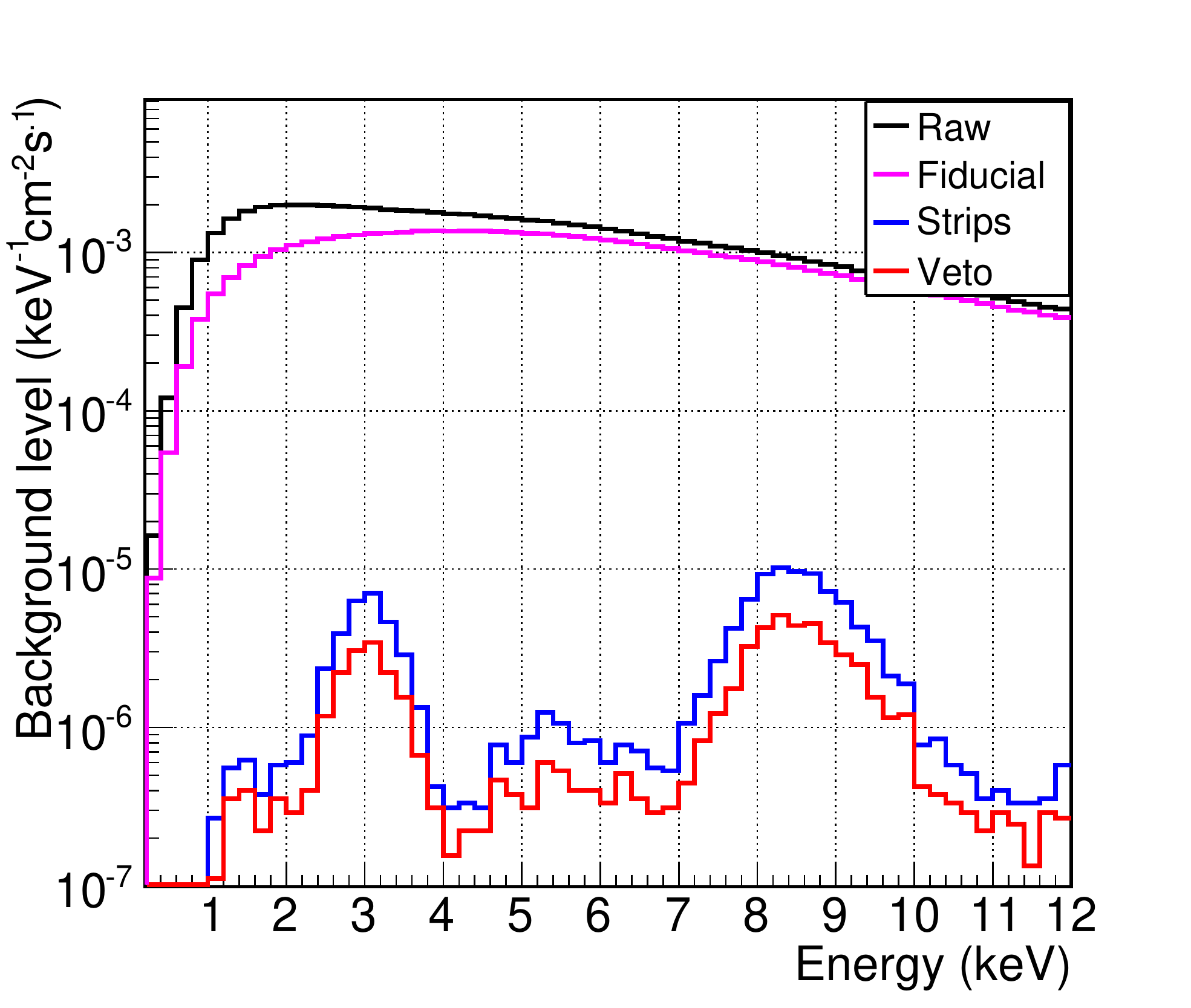}
\includegraphics[height= 6cm]{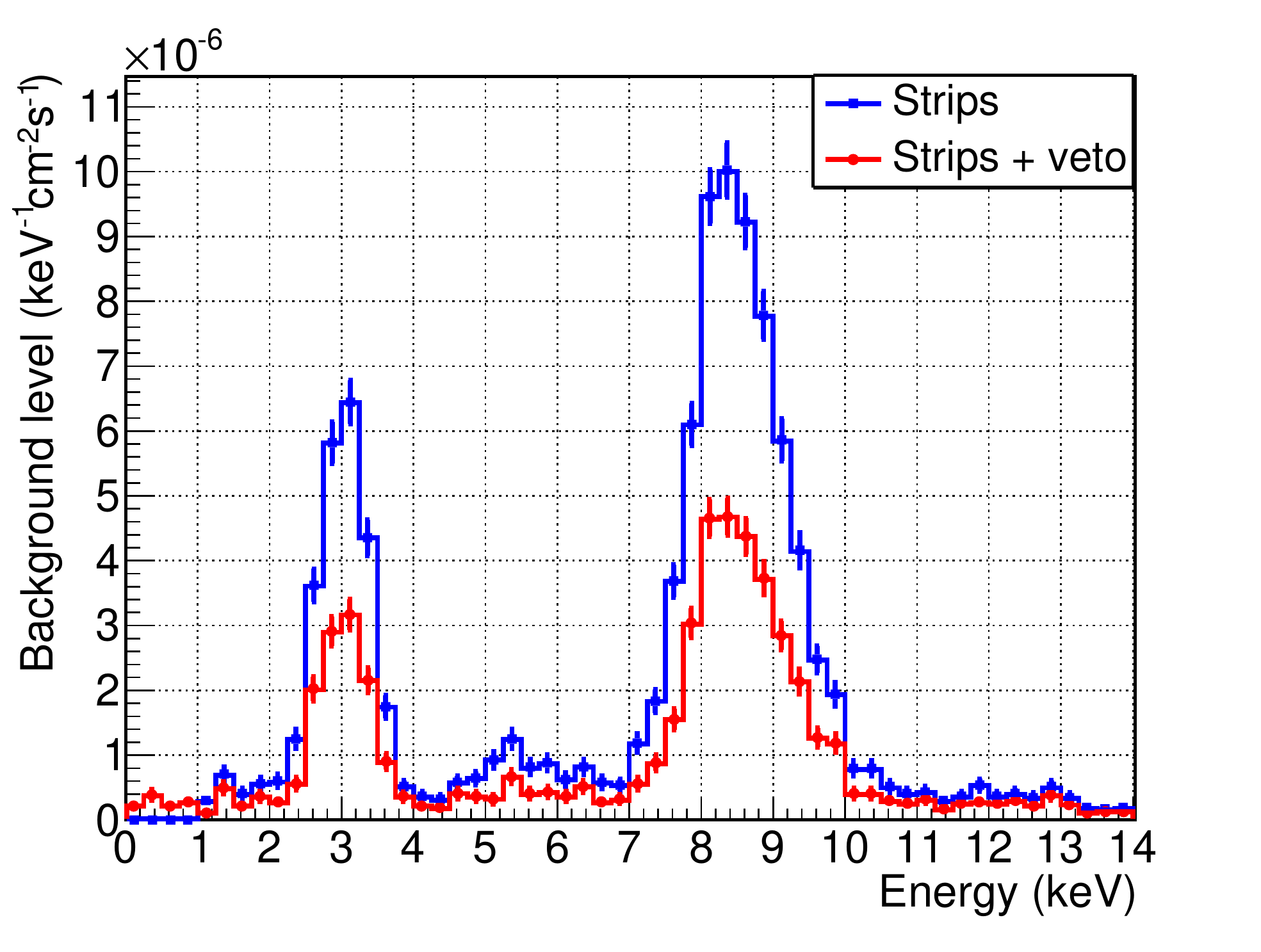}
\caption{Left: Background energy spectra of the sunrise CAST detector in the signal area, both before and after off-line cuts. Remaining background without (blue) and with (red) the veto coincidence criterion are plotted. The integrated [2--7]~keV levels of both spectra correspond to $(1.63 \pm 0.04) \times~10^{-6}$~\ckcs, and $(0.83 \pm 0.03) \times~10^{-6}$~\ckcs respectively. Right: Background after off-line cuts in linear scale without (blue) and with (red) the veto coincidence; the peak at 8~keV is due to copper fluorescence from the materials at the entrance pipe, and the readout while the 3.2~keV one corresponds to the argon fluorescence.}
\label{fig:bkgspectra}
\end{figure}

The actual implementation of these techniques must take into account the intrinsic rejection capabilities of the detector, provided by the highly granular Micromegas readouts. For example, muons directly crossing the detector are easily distinguished from signal (point-like) events. Yet secondary radiation (e.g. x-ray fluorescence), produced by muons in the materials close to the detector sensitive volume would eventually contribute to background. Only in the last generation of CAST detectors, with background levels at the few 10$^{-6}$~\ckcs, muon tagging has a significant impact. Off-line discrimination using the topological information currently allows for a reduction of the raw background level of about a factor $\sim$10$^2$--10$^3$. The final background levels currently achieved in CAST detectors are at around 10$^{-6}$~\ckcs~\cite{Aune:2013nza} in the [2--7]~keV energy range. The best result is obtained by the sunrise detector, with a value of operation in CAST of as low as (0.83 $\pm$ 0.03)~keV$^{-1}$cm$^{-2}$s$^{-1}$ integrated over the entire energy range and detector area. The background energy spectrum of this detector is shown in the Figure~\ref{fig:bkgspectra} with and without the veto coincidence rejection. It is to be noted that 8~keV fluorescence (from copper K$_{\alpha}$ emission), its escape peak at 5~keV as well as the argon K$_{\alpha}$ fluorescence line at 3~keV dominate the spectral distribution. The flat background outside the fluorescence peaks is probably of at least a factor 2 lower. It is worth mentioning that the background is not completely homogeneous over the detector area, with a slight increase at its center, where the signal spot is expected. This raises the effective background level at the signal spot to about (1.10 $\pm$ 0.08)~keV$^{-1}$cm$^{-2}$s$^{-1}$. This effect may be related to the opening angle of the pipe towards the magnet.


Summarizing, the above values correspond to 0.13 background counts per day in the energy region-of-interest and signal spot area. Given that the sunrise system spends only $\sim$1.5 hours tracking the Sun per day, it takes CAST about four months to have an average of $\sim1$ count of background in this detector in present conditions.These numbers represent the state-of-the-art in low-background x-ray detection, being the lowest detector background ever obtained in a system exposed to external 1--10 keV x-rays, and it is the main experimental result so far of the T-REX R\&D line for axion research. However, as part of the R\&D to understand the origin of the remaining background levels, a replica of the detectors above described has been installed in an underground location at the LSC~\cite{Aune:2013nza}. The background achieved with this detector is as low as $\sim$10$^{-7}$~keV$^{-1}$cm$^{-2}$s$^{-1}$. This result suggests that the background limitation at surface be still of cosmic origin, and is important to establish the future steps to further reduce the background levels and eventually reach the very demanding specifications set for IAXO. The interpretation of these results and the strategy for IAXO is described in the next section.



\section{Prospects for IAXO}
\label{sec:iaxo}

\begin{figure}[!t]
\centering
\includegraphics[width=0.75\textwidth]{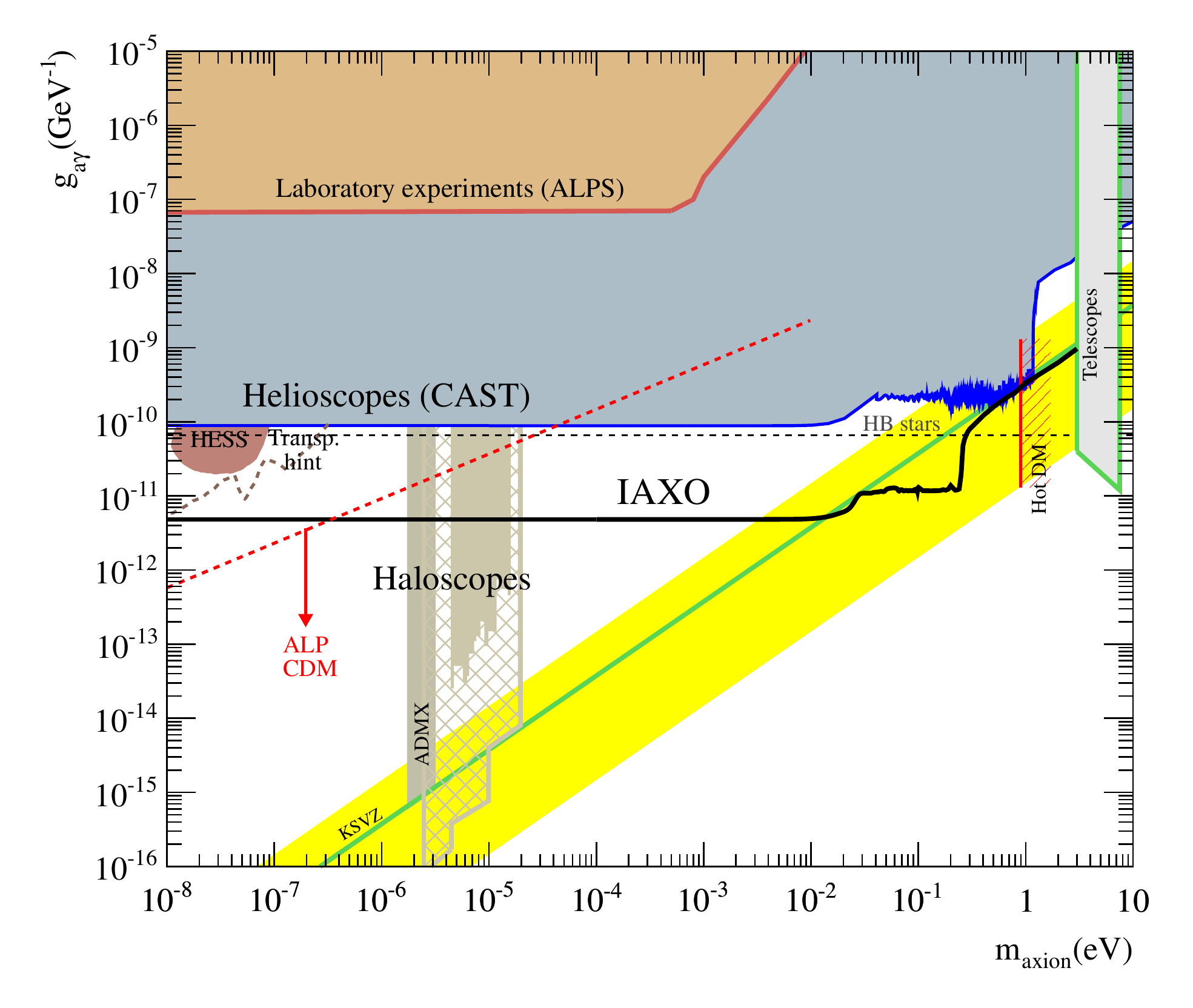}
\caption{ Expected sensitivity of IAXO (black solid line) in the axion and axion-like particle parameter space, compared with current helioscope bounds from CAST. The yellow band represent the approximate theoretical extent of axion model parameters. Other lines and areas of interest are also shown. We refer to \protect\cite{Irastorza:1567109} for a detailed explanation of the regions and bounds of the plot.}
\label{fig:iaxoLimit}
\end{figure}

The developments described in previous sections have a strong motivation as part of the preparatory activities for IAXO, the future next-generation axion helioscope now under proposal~\cite{Irastorza:2011gs,Armengaud:2014gea}. IAXO envisions the construction of a large superconducting 8-coil 20~m long toroidal magnet optimized for axion research~\cite{Shilon:2012te}. In between the coils, the IAXO magnet will host eight 60~cm diameter bores, each of them equipped with x-ray optics~\cite{doi:10.1117/12.2024476} focusing the signal photons into $\sim$0.2~cm$^2$ spots that are imaged by ultra-low background x-ray detectors. Microbulk Micromegas like those used in CAST are one of the detection technologies candidates to be used in IAXO. The magnet will be built into a structure with elevation and azimuth drives that will allow solar tracking for $\sim$12~hours each day. In terms of signal-to-noise ratio, IAXO will be about 4--5 orders of magnitude more sensitive to Primakoff solar axions than CAST, which translates into a factor of $\sim$20 in terms of the axion-photon coupling constant $\gagamma$. That is, this instrument will reach the few $\times$10$^{-12}~{\rm GeV}^{-1}$ regime for a wide range of axion masses up to about 0.25~eV (see Figure~\ref{fig:iaxoLimit}).

%

Although the developments described in the previous sections have led to a reduction of the detector backgrounds allowing for CAST to almost reach effectively the zero-background situation, and thus its ultimate sensitivity to solar axions, the higher detector exposure foreseen for IAXO would benefit from even lower background levels. Indeed, among the general IAXO specifications drawn in the conceptual design~\cite{Armengaud:2014gea}, the required detector background level is set between $10^{-7}-10^{-8}$~\ckcs, 1--2 orders of magnitude lower than the levels achieved by the latest CAST detectors, albeit only slightly below the levels seen in the underground setup. Just for the purpose of illustration, we note that a background of $10^{-8}$~\ckcs corresponds to about $\sim$5 background counts in the signal spot areas and energy RoI of IAXO during a 3-year data taking campaign. In the following we discuss the prospects to reach such background levels and the steps being carried out at the moment to demonstrate them experimentally.

Despite the efforts taken with the CAST detectors, the shielding capabilities are still somehow limited by mechanical and space constraints inherent to the CAST experiment. This is primarily linked to the relatively small room available at the extremes of the magnet, to the weight limits of the existing structure to support detectors and shielding as well as to their combined movement with the magnet during the Sun tracking. The fact that the underground replica of the detector has obtained a level of $10^{-7}$~\ckcs~\cite{Aune:2013pna} clearly indicates that background at surface is still limited by cosmic-related sources. As part of the IAXO Technical Design Report, a prototype of x-ray detector, called IAXO-D0, is being built to experimentally prove this hypothesis.

\begin{figure}[!t]
\centering
\includegraphics[width=0.8\textwidth]{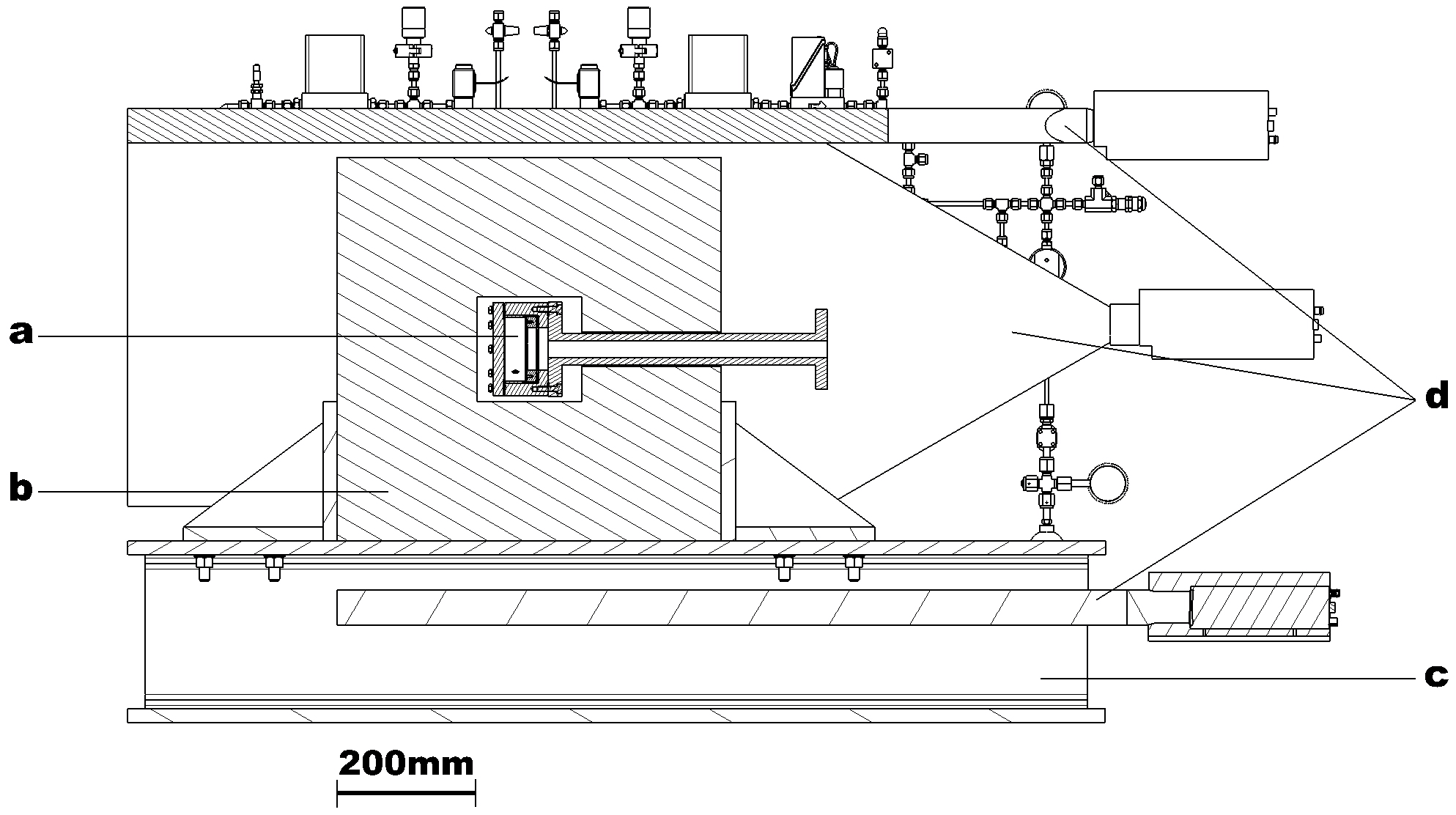}
\caption{ Sketch of the IAXO-D0 design, with some of the different parts of the set-up labelled: detector chamber (\textbf{a}), 20~cm thick lead shielding (\textbf{b}),
supporting structure (\textbf{c}) and plastic scintillators (\textbf{d}).}
\label{fig:iaxod0}
\end{figure}

IAXO-D0 is conceived as an evolution of the latest sunrise CAST detector, without, however, the constraints imposed by CAST geometry. The passive shielding will be enlarged to --at least-- 200~mm of lead in all directions, while the active shielding will be improved to the point of an effective $4\pi$ coverage of cosmic muons. The setup design is shown in Figure~\ref{fig:iaxod0}. It should allow to reject cosmic-related background to levels at least comparable to the current result underground (where the muon component is negligible). To go to even lower background levels, intrinsic radioactivity sources must be studied. While the radiopurity of all detector components has already been exhaustively controlled and reduced to levels corresponding to negligible contributions to the detector background (according to current background model simulations), the $\beta$ emitter $^{39}$Ar isotope present in the Ar gas used in the chamber may remain now an important component. Indeed, estimations show that it could even be a dominating background component at the $10^{-7}$~\ckcs level. Therefore, one of the design goals of IAXO-D0 will be to test operation with different gases in closed recirculation mode. One option is to use similar Ar-based mixture but depleted from the $^{39}$Ar isotope. Depleted Ar has been obtained from underground gas sources in the context of liquid Ar DM detectors~\cite{Agnes:2015ftt}. Another option under consideration is the use of a Xe mixture. The experience obtained in the T-REX double beta decay R\&D work described in~\cite{trexbbreview} may be highly useful for this purpose. This latter option would also be appealing in order to suppress the escape and the Ar fluorescence peaks that dominate the background spectra (see Figure~\ref{fig:bkgspectra} right), considering that the Xe fluorescences in this energy range are highly suppressed.

\begin{figure}[!t]
\centering
\includegraphics[width=0.6\textwidth]{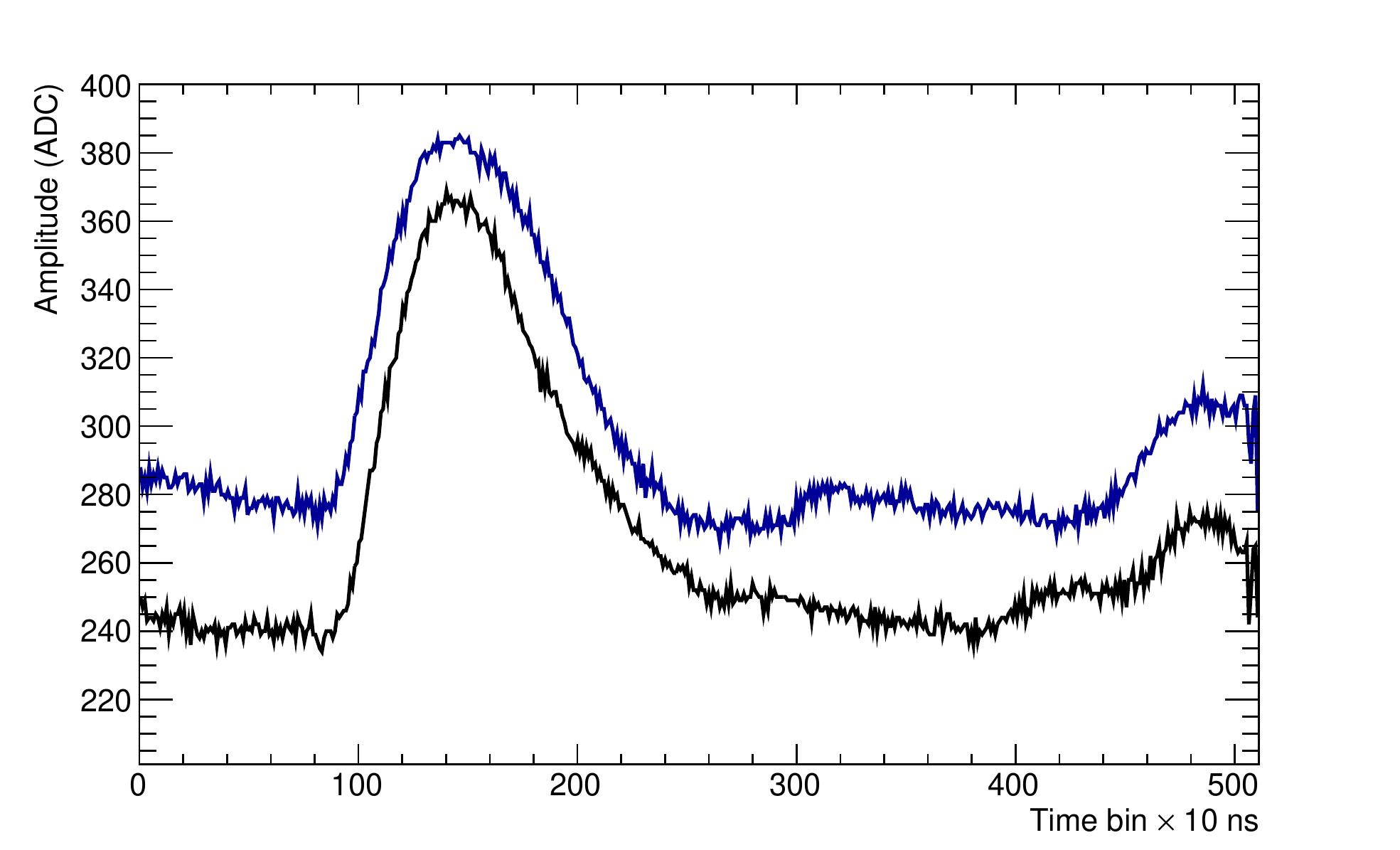}
\caption{ Example of a low energy ($\sim$300 eV) event, producing pulses in two strips, acquired during a test run with a CAST detector equipped with AGET electronics, in non-optimal conditions of detector noise. The signal to noise ratio observed suggests that thresholds close to $\sim$100 eV are achievable.}
\label{fig:threshold}
\end{figure}

Finally, IAXO-D0 will be equipped with AGET-based front-end electronics~\cite{Baron:2011} with autotrigger capability for every single readout channel. This will allow to substantially improve the energy threshold of the detector. Current DAQ implementation use the mesh signal to trigger the strip electronics. The energy threshold is thus limited by the electronic noise of the mesh electrode, of relatively high capacitance, to values around $\sim$1~keV. The new electronics should allow to trigger on the strip signals, that are about 100 times less capacitive. Figure~\ref{fig:threshold} shows a pulse of $\sim$400 eV acquired with a CAST detector equipped with AGET electronics. Judging by the signal-to-noise ratio of such a pulse, energy thresholds of down to 100~eV seem achievable. This feature is of interest to IAXO, in order to gain sensitivity to additional solar axion and axion-like particle emission channels, e.g. to solar axions generated by non-hadronic processes~\cite{Redondo:2013wwa}.

Taking into account all of the above considerations, the microbulk Micromegas technique has solid prospects to obtain the ultra-low levels of background required for IAXO, that still lies 1--2 orders of magnitude below the experimental levels demonstrated in CAST. The experimental demonstration of this fact is the main goal of IAXO-D0. Even in the case that these prospects are not realized in a first stage, the operation and results with the IAXO-D0 prototype will be instrumental to design the next steps in the quest to achieve that goal.

%

\section{Gas TPCs for low-mass WIMPs}
\label{sec:wimps}

As mentioned in the introduction, some recent experimental and phenomenological efforts have been focused on the study of WIMPs in the low mass range (i.e. $M_W < 10-20$~GeV). The interest in this region of the parameter space, traditionally out of reach of mainstream experiments, was increased by the appearance of a number of hints that could be interpreted as due to low-mass WIMPs~\cite{Aalseth:2011wp,cresstII} (although those interpretations seem to have weakened over time~\cite{Agnese:2014aze,Angloher:2014myn}). In addition, the well-known and persistent DAMA claim~\cite{refId0}, despite having been excluded by many other experiments, might be reconciled only within very non-standard model assumptions, some of them invoking low-mass WIMPs~\cite{Savage:2008er}. Independently of the weight one gives to those hints, or to the theoretical motivation of low-mass WIMPs, it is clear that in the present situation of WIMP searches, as much as it is important to extend the current generic ($\sigma_N , M_W$) sensitivity frontline to lower $\sigma_N$ values, so it is extending it to lower $M_W$ values too.

If the DM halo is made of such low-mass WIMPs, conventional experiments may have some difficulties to detect them. The nuclear recoils produced by these WIMPs in the heavy target nuclei that are most popular (e.g. Ge or Xe) typically fall in energies that lie below the energy threshold set by the electron-nuclear recoil discrimination mechanism. Even if typical sensitivity projections give the impression (see e.g. sensitivity plot in Figure~\ref{fig:exclusion}) that long enough exposure may compensate this and provide competitive sensitivity at low masses, this needs to be taken with a lot of caution. The lower mass edge of the sensitivity lines derived by mainstream experiments typically rely on a very small (1\% or lower) fraction of the WIMP interactions in the detector, those corresponding to the high velocity tail of the distribution (with kinetic energies enough to produce a nuclear recoil visible in the detector). But precisely this part of the distribution is the most uncertain, and in some plausible galactic halo models it can altogether disappear (e.g. those with lower maximum WIMP velocity).

In order to have a substantial fraction (order 50\%) of the WIMP spectrum over the experimental threshold, the use of light target nuclei, as well as techniques with intrinsically low energy detection threshold are needed. These requirements are incompatible with the discrimination between nuclear and electron recoils, whose observable features become blurred at low energies.
It is clear that a new category of experiments, optimized in this way, are needed to access the low mass WIMP region. Some experimental efforts (still at modest scales) are already being carried out in this direction (e.g.~\cite{Chavarria:2014ika,Agnese:2015nto,PhysRevD.90.091701} among others). As the background levels in these experiments must rely on more conventional handles like e.g. ultra-high levels of radiopurity of the detector components, the scale of these experiments remains so far  relatively modest (still below the kg level of target mass).

The use of gas TPCs with Micromegas readouts has been recently proposed~\cite{Iguaz:2015myh} to search for low mass WIMPs, as part of the T-REX project. Many of the technical advantages exploited in the development of the detectors for axion research are of direct application also in this case. Namely, the possibility to build Micromegas readouts with radiopure materials~\cite{Cebrian:2013mza} and with a signal extraction scheme of extreme radiopurity (see discussion in section 3 of the companion paper~\cite{trexbbreview}); or the capability to use off-line topological discrimination techniques based on the highly granular readout. In addition, the way event detection happens in gas (i.e. drift of charge and signal amplification confined in the Micromegas structure) allows, in principle, to reach very low energy threshold even in relatively large size detectors. Another aspect, very important for application to WIMP searches, are the scaling-up prospects. Technical solutions for scaling-up via tessellation of identical microbulk have been defined in the section 7 of~\cite{trexbbreview}.

In order to experimentally explore this concept, the TREX-DM prototype has been built. Many of the technical specifications affecting the Micromegas readouts are inherited from the axion development presented in the previous sections. TREX-DM has been designed, however, putting the focus on testing the specifics of the application to WIMP searches. Most particularly, TREX-DM will host about 1000 times more target mass than the CAST/IAXO prototypes, and it is conceived to be operated underground and adequately shielded. The prototype is described in the next section, along with the first results from its commissioning. As shown in section~\ref{sec:prospects}, TREX-DM could already produce interesting physics results, depending on the  final background to be achieved underground.

\section{The TREX-DM prototype}
\label{sec:trexdm}

\begin{figure}[ht!]
\centering
\includegraphics[width=0.8\textwidth]{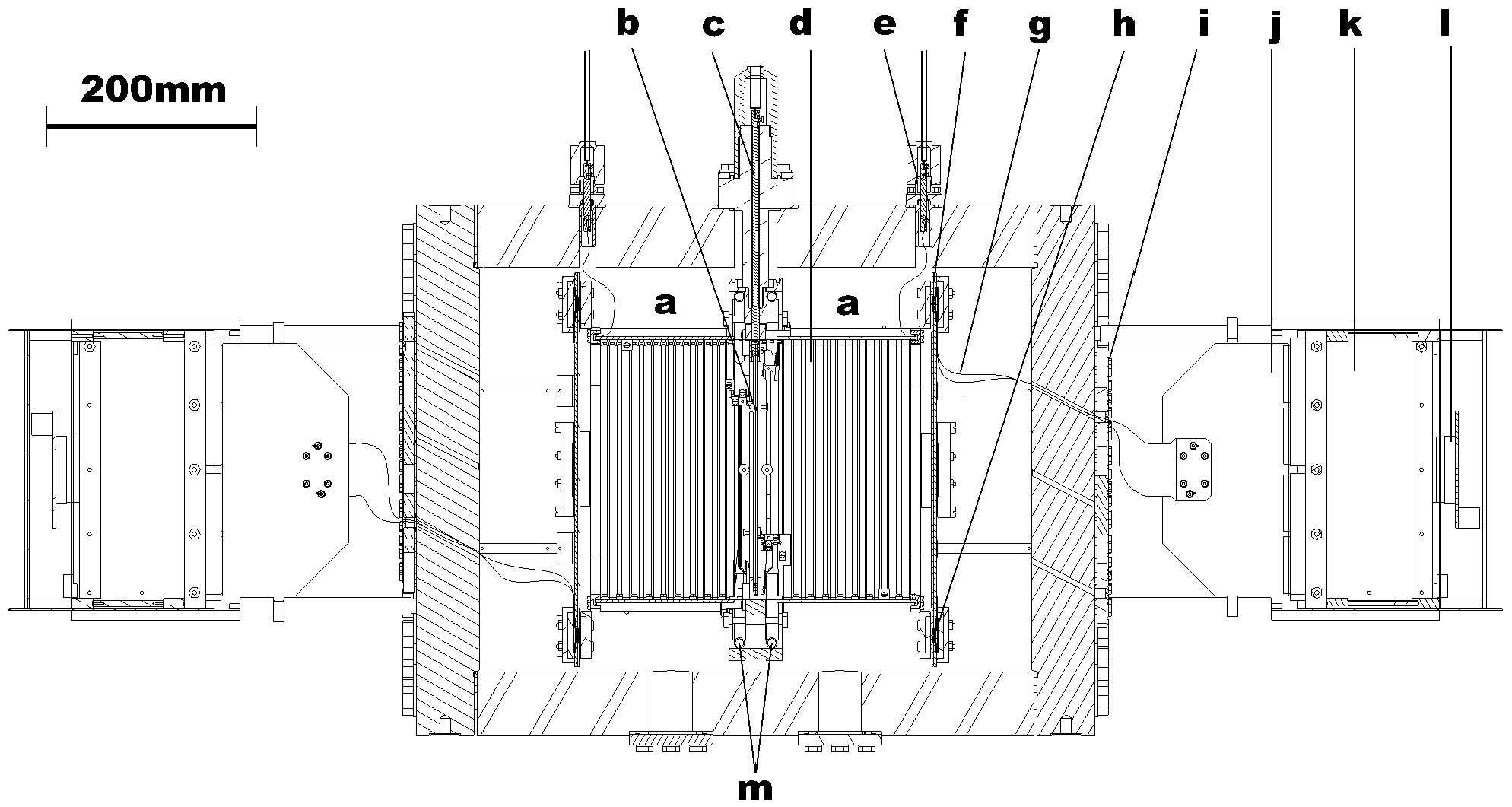}
\caption{Sketch of the experimental setup extracted from \cite{Iguaz:2015myh}.
The different components are described in detail in the text:
active volumes (a), central cathode (b), high voltage feedthrough (c), field cage (d), \emph{last ring} feedthrough (e),
Micromegas readout planes (f), flat cable (g), Samtec connectors (h), signal feedthroughs (i), interface card (j),
AFTER-based front-end cards (k) and front-end mezzanine boards (l), and calibration tube (m).}
\label{fig:TREXDMSchema}
\end{figure}

\begin{figure}[ht!]
\centering
\includegraphics[width=48mm]{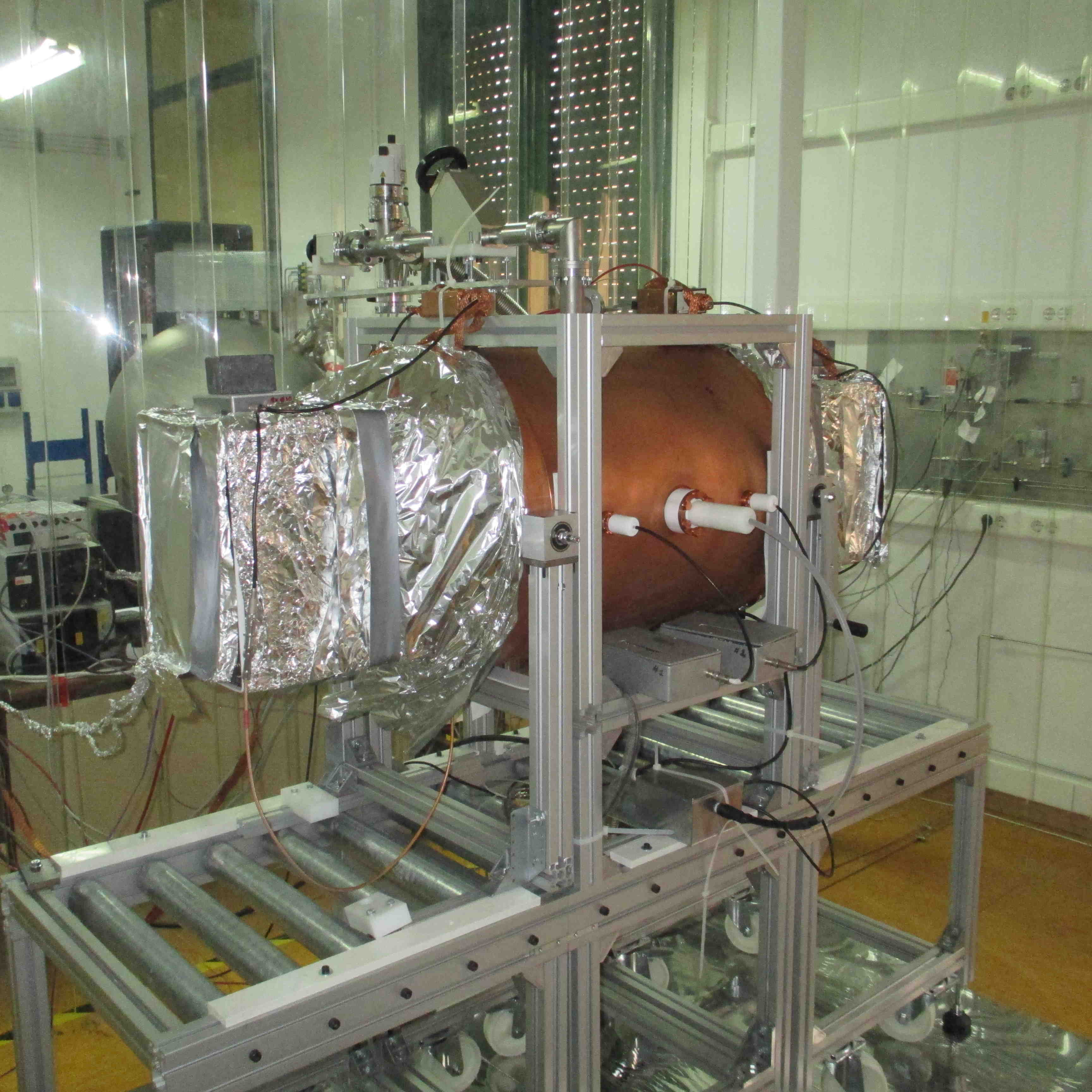}
\includegraphics[width=48mm]{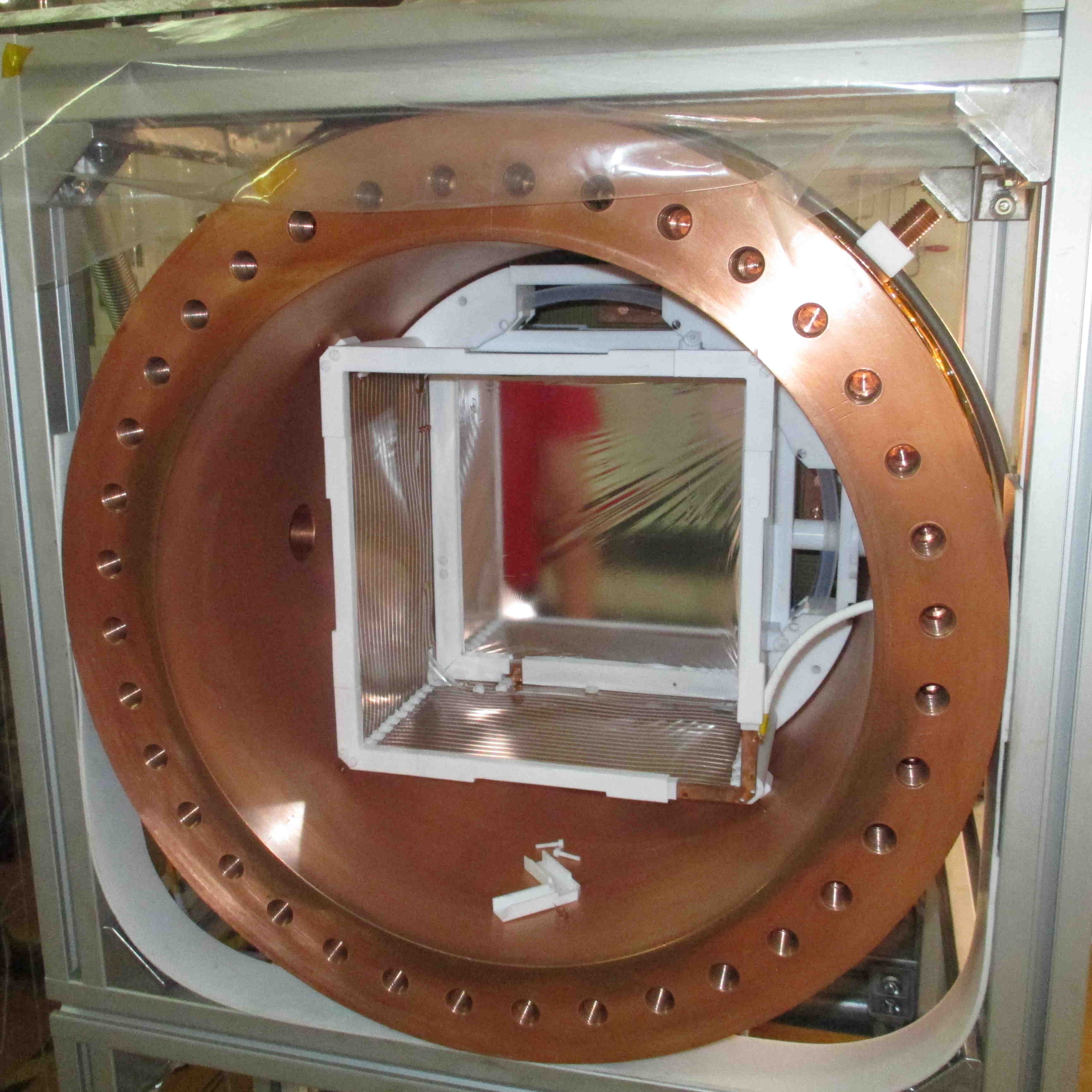}
\includegraphics[width=48mm]{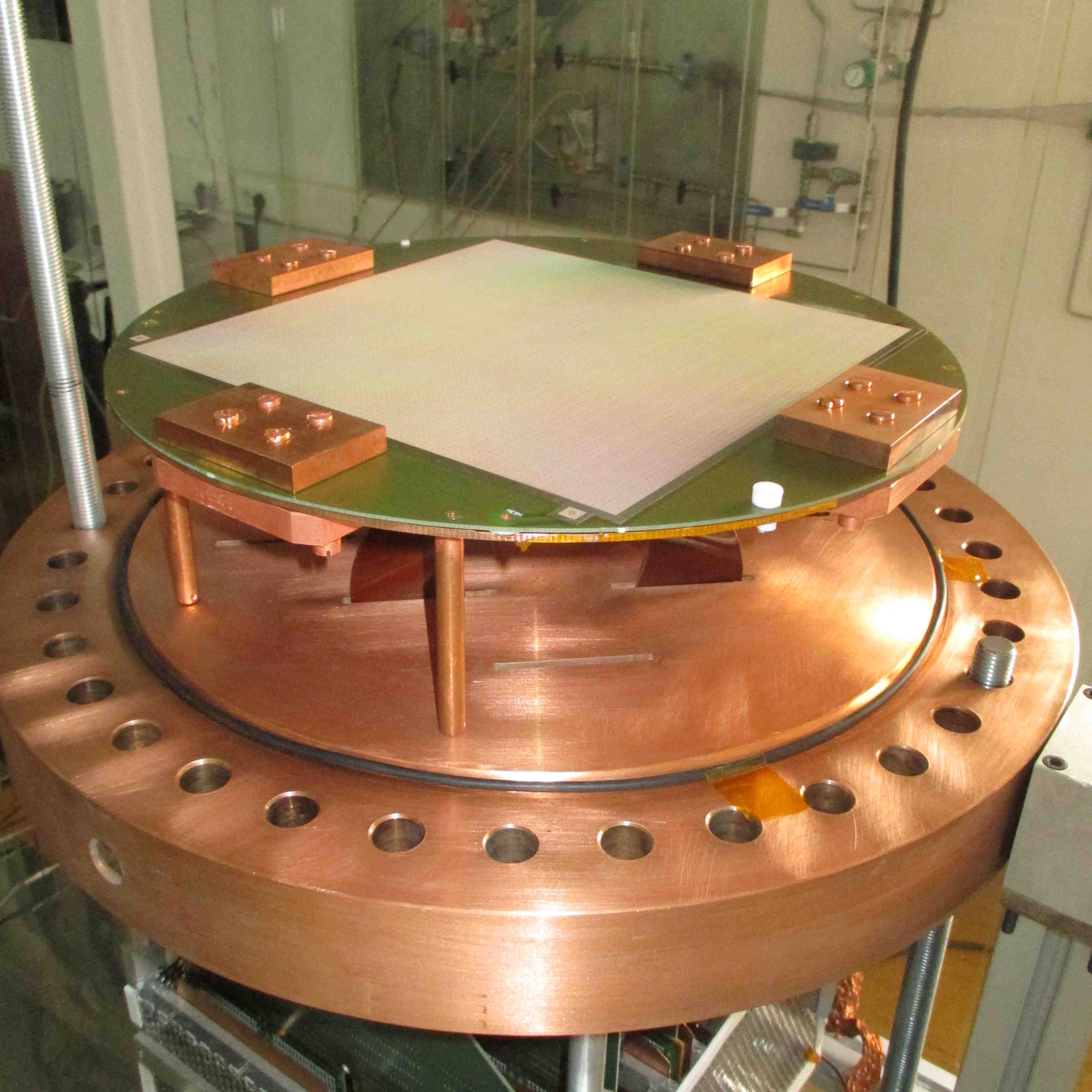}
\caption{Left: a view of the TREX-DM experiment during its commissioning.
Center: the inner part of the vessel, when one of the flat end caps has been removed. One can see the mylar foil
used as central cathode, the copper-kapton printed circuit used as field cage
and the copper squared ring, covered by a teflon gasket, where the drift chain ends.
Right: a Micromegas readout plane installed at its support base, which is screwed to one of the vessel's caps.
Four flat cables extract strips signals generated at the readout plane outside the vessel
by four feedthroughs and are then connected to four interface cards, and subsequently to four FEC cards.}
\label{fig:TREXDMSetup}
\end{figure}

The TREX-DM TPC has been designed to host 0.3~kg of argon mass at 10~bar (or, alternatively, 0.16~kg of neon). The design of the detector is shown in Figure~\ref{fig:TREXDMSchema} and a picture of the overall setup during its commissioning is shown on the left of Figure~\ref{fig:TREXDMSetup}. It is composed of a cylindrical vessel made of radiopure copper, with an inner diameter of 0.5~m, a length of 0.5~m and a wall thickness of 6~cm. These dimensions are set by the requirements that the vessel holds up to 10~bar of pressure, while at the same time constitutes the innermost part of the shielding. The vessel is divided into two active volumes ($a$ in Figure~\ref{fig:TREXDMSchema}) by a central mylar cathode ($b$), which is connected to high voltage by a tailor-made feedthrough ($c$). At each side there is a 19~cm long field cage ($d$ in Figure~\ref{fig:TREXDMSchema} and Figure~\ref{fig:TREXDMSetup}, center), defined by a series of copper strips imprinted on a kapton substrate supported by four teflon walls. The copper strips are electrically interconnected by 10~M$\Omega$ resistor chain that ends at a copper squared ring ($e$), connected to a customized high voltage feedthrough. Its voltage is adjusted by an external variable resistor, in order to get a uniform field in the active volume.

At each of the vessel end caps, the Micromegas readout plane (\textbf{f}) in Figure~\ref{fig:TREXDMSchema} and Figure~\ref{fig:TREXDMSetup} (right) is supported by a copper base, which is then attached to the cap via four copper columns. The readout pattern is based on the CAST Micromegas readouts~\cite{Aune:2013pna}, extended over an active surface of $25.2 \times 25.2$~cm$^2$. The current readout implementation, just for the commissioning phase, follows the \emph{bulk} technology~\cite{Giomataris:2004aa}, but a radiopure microbulk Micromegas version will be installed for the definitive operation underground. Strip signals are extracted from the vessel via four flat cables (\textbf{g}). Each flat cable comes out from the vessel through a slit at one of the vessel caps and a feedthrough screwed to the external face of the end cap~(\textbf{i}). The flat cable is fixed to the feedthrough by a teflon gasket which is then glued by radiopure epoxy. One of the cable ends is linked to one of the readout footprints by a 300-pin Samtec connector (\textbf{h}), and the other one connected to an interface card that routes the signals to an AFTER-based front-end card~\cite{Baron:2008zza,Baron:2011}~(\textbf{j}).

The detector can be calibrated with a $^{109}$Cd source (x-rays of 22.1~keV and 24.9~keV) that is placed at the tip of a plastic cable that can slide inside a teflon tube (\textbf{m}) that enters the vessel through a leak-tight port. In this way the source can be positioned at four different points per active volume, situated at the corners of the central cathode, to obtain an adequate illumination of the full detector volume. Finally, two gas ports, at the bottom and the top of the vessel, allow for gas circulation. The pressure and flow of the gas are adjusted by corresponding pressure transducer and mass flowmeters via a slow control.

The detector has been commissioned on surface equipped with bulk Micromegas readouts and AFTER-based data acquisition. The Micromegas readout planes have been characterized in Ar+2\%iC$_4$H$_{10}$ for pressures between 1.2~bar and 10~bar, in steps of 1~bar with the Cd source. They show a wide range of drift fields for which the mesh is transparent to primary electrons, similar to those of other bulk detectors~\cite{Iguaz:2011yc}. Both readouts show a similar gain for the same mesh voltage, their maximum operation gain decreasing with the gas pressure, from $3\times10^3$ at 1.2~bar down to $5\times10^2$ at 10~bar. The energy resolution for the 22.1~keV peak has been measured to be from 16\%~FWHM at 1.2~bar to 25\%~FWHM at 10~bar, although better values are expected for mixtures with higher amount of quencher. The signal-to-noise ratio in the high-capacitance mesh signal allows to see events of energies of 1.0~keV at 1.2~bar and 1.4~keV at 3~bar, only slightly worse than the values obtained with smaller Micromegas detectors~\cite{Aune:2013pna}. These values reassure us that an effective threshold of down to 100~eV (see discussion in section~\ref{sec:iaxo} and Figure~\ref{fig:threshold}) will be reachable once the final AGET-based front-end electronics~\cite{Baron:2011}, able to trigger on the ($\sim$100 times less capacitive) strip signals, will be implemented.

\section{Prospects for the search of low-mass WIMPs}
\label{sec:prospects}

In order to have an estimation of the achievable sensitivity of TREX-DM to low-mass WIMPs, a preliminary background model of the experiment has been built assuming operation at the LSC. Two options for the TPC gas have been considered: Ar+2\%iC$_4$H$_{10}$ and Ne+2\%iC$_4$H$_{10}$, both good candidates as targets for this purpose. At 10~bar, they correspond to a total target mass of 0.30~kg and 0.16~kg respectively. The model is fed with data from a dedicated screening program including every material used in the detector construction: the vessel, the field cage, the shielding and the readout planes. The screening is carried out with germanium gamma-ray spectrometry at the LSC~\cite{Cebrian:2010ta,1748-0221-8-11-C11012} and, complementing these results, with Glow Discharge Mass Spectrometry (GDMS). Details on the measurements have been presented in the companion paper, and in more detail in~\cite{Iguaz:2015myh}. The model includes a simulation of the detector response based on a Geant4 program~\cite{Agostinelli:2002hh} and the REST code~\cite{ATomas_PhD}. The latter simulates the signal processes inherent to TPCs, from the creation of primary electrons until the generation of signals at mesh and strips.

The radioactive isotopes of the most relevant components of the prototype have been simulated, and the results have been scaled by the activities of the screening program or, for some of the components, values were drawn from the literature. Finally, for the specific case of argon-based mixtures and the $^{39}$Ar isotope, the activity in~\cite{Agnes:2015ftt} has been used. External sources of background, like environmental gammas or neutrons, have not been included in this first version of the model, under the assumption that the detector will be surrounded by shielding, effective enough to bring those contributions to negligible levels.

Simulated events are filtered using criteria defined to discriminate low energy signal (i.e. point-like) events from some of the more complex background topologies like e.g. that of muons. The signal pattern features are characterized using two alternative simulated event populations, calibration x-rays from a $^{109}$Cd source or neutron events from a $^{252}$Cf source. While the second analysis is supposed to be closer to reality (as neutrons, like WIMPs, produce nuclear recoils) the first analysis is useful to make comparison with real data (until experimental neutron calibrations are available). In any case, at high pressure both low energy x-ray and neutron events show very similar topologies, and indeed both analysis yield similar results. We refer to ~\cite{Iguaz:2015myh} for a more detailed explanation of the background model.

\begin{figure}[ht!]
\centering
\includegraphics[width=100mm]{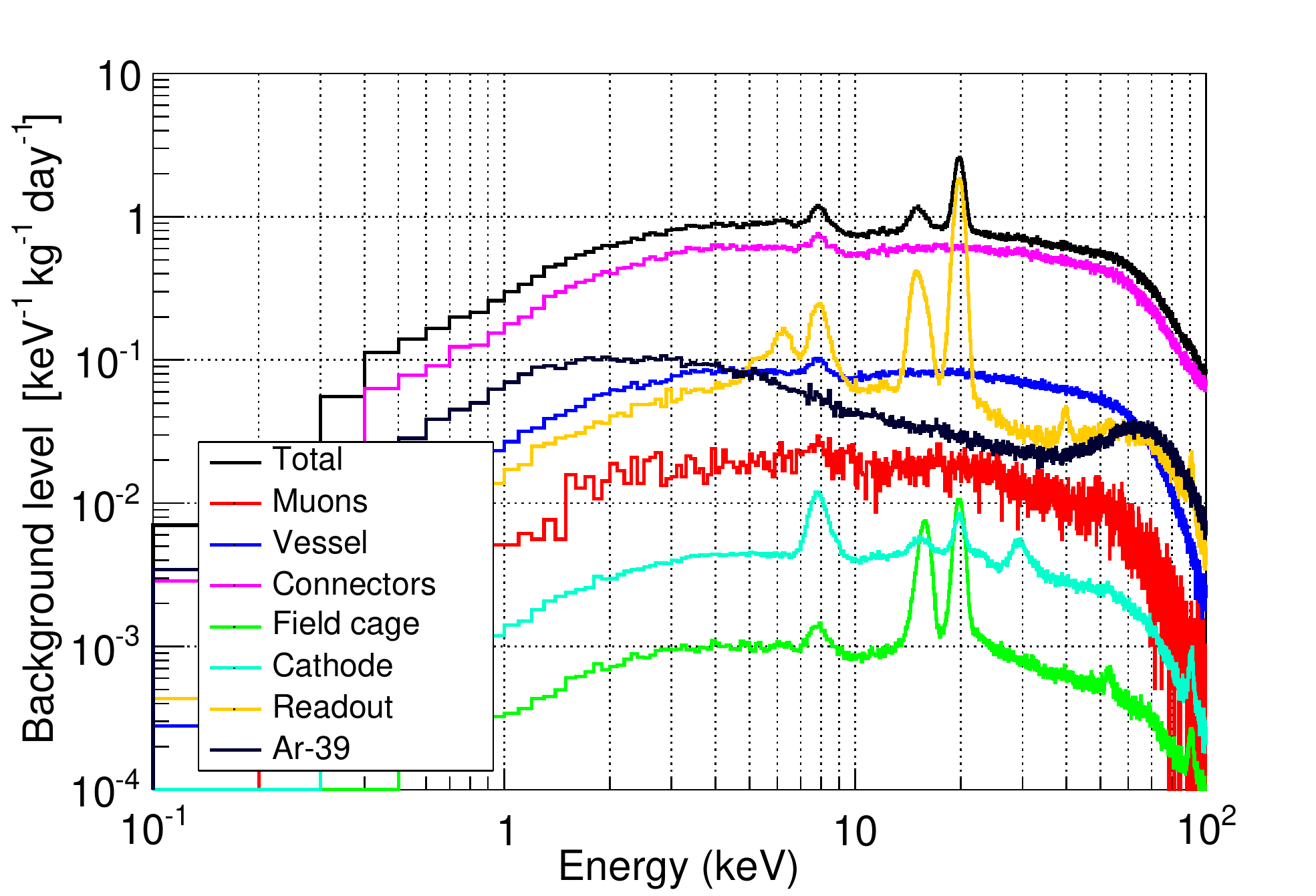}
\caption{Background spectrum expected in TREX-DM experiment (black line) if operated in Ar+2\%iC$_4$H$_{10}$ at 10 bar according to the preliminary background model referred to in the text.
The contribution of the different simulated components is also plotted: external muon flux (red line),
vessel contamination (blue line), connectors (magenta line), field cage (green line), central cathode (brown line),
Micromegas readout planes (orange line) and $^{39}$Ar (dark blue line).
Further details can be found in~\cite{Iguaz:2015myh}.}
\label{fig:TREXDMBackSpec}
\end{figure}

The final background level predicted by the model, fixing a signal efficiency of 80\% for the off-line criteria, suggest that a very competitive $\sim$1 keV$^{-1}$ kg$^{-1}$ day$^{-1}$ in the 1--10~keV energy range is achievable both for argon and neon-based mixtures. The background level decreases for sub-keV energies, however the validity of the particle interaction models at these energies is probably questionable. The total spectral distribution, as well as that of each of the simulated components, are shown in Figure~\ref{fig:TREXDMBackSpec}. The background is dominated by the radioactivity of the connectors, which in future versions of the detector could easily be replaced by cleaner ones; or it could be placed farther from the sensitive volume and better shielded. The contributions from the readout plane and vessel (that come however from upper limits to their radioactivity) lie at a level of one order of magnitude lower. These considerations suggest that there is no intrinsic limitation in this detection concept to reach levels down to, or even lower than, $\sim$0.1 keV$^{-1}$ kg$^{-1}$ day$^{-1}$. These results, although encouraging, must await experimental validation by actual data taking of TREX-DM underground. The experimental study of background limitations at these energies at the radiopurity levels stated is \emph{per se} an experimental question of high interest.

\begin{figure}[!hbt]
\centering
\includegraphics[width=0.5\textwidth]{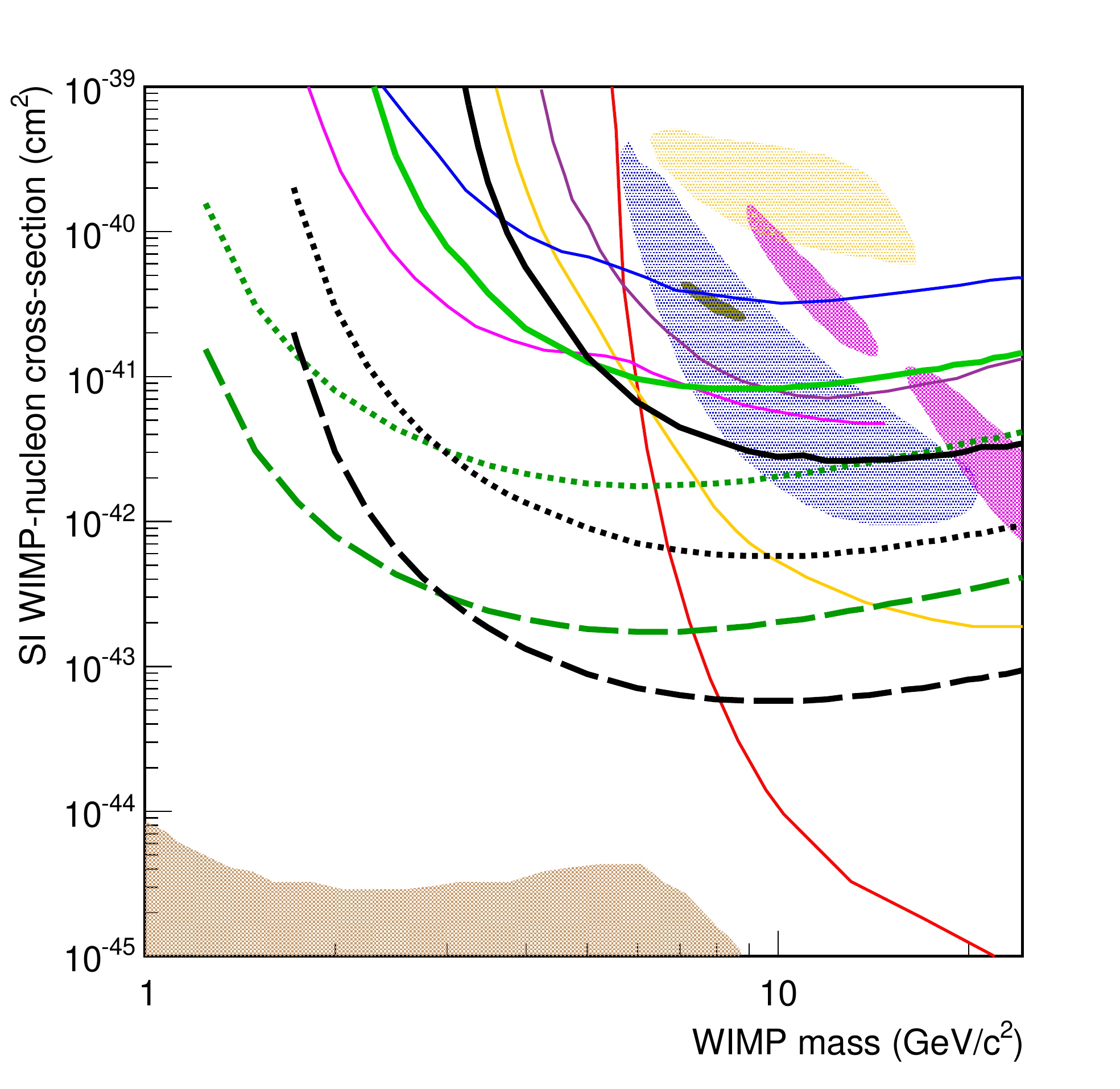}
\caption{The projected sensitivity of TREX-DM (at 90\% confidence level) assuming an exposure of 1~kg$\cdot$y in argon (\emph{black thick lines}) and neon (\emph{green thick lines}) with a conservative (\emph{solid}) and realistic (\emph{dotted}) assumptions on the background levels of 10 and 1~keV$^{-1}$Kg$^{-1}$d$^{-1}$, respectively, and an energy threshold of 0.4 keVee and 0.1~keVee, respectively. The potential sensitivity of a scaled-up detector with a threshold of 0.1~keVee, background of 0.1~keV$^{-1}$kg$^{-1}$d$^{-1}$ and an exposure of 10~kg$\cdot$y is also shown (\emph{dashed lines}). Closed contours correspond to the positive interpretations of the excess counts of CDMS II Si~\cite{PhysRevLett.111.251301} (\emph{blue}, 90\% C.L.), CoGeNT~\cite{PhysRevD.88.012002} (\emph{dark gray}, 90\% C.L.), CRESST-II~\cite{cresstII} (\emph{magenta}, 95\% C.L.), and DAMA/LIBRA~\cite{refId0,Savage:2008er} (\emph{tan}, 90\% C.L.). For comparison we also show 90\% C.L. exclusion limits from SuperCDMS~\cite{Agnese:2014aze} (orange), CDMSlite~\cite{Agnese:2015nto} (\emph{magenta}), LUX~\cite{PhysRevLett.112.091303} (\emph{red}), and CDEX1~\cite{PhysRevD.90.091701} (purple) and CRESST-II 2015~\cite{Angloher:2015ewa} (\emph{blue}). The brown shaded region corresponds to the sensititivy limit imposed by the solar neutrino coherent scattering background~\cite{Billard:2013qya}.}
\label{fig:exclusion}
\end{figure}

Assuming that the background levels predicted by the model are experimentally realized, we have computed the projected sensitivity that TREX-DM could achieve to low-mass WIMPs. A binned Poisson method with background subtraction and energy binning of 100~eVee has been used. For the signal calculations, the Lindhard parametrization for the quenching factor has been used, as well as a standard WIMP halo model with Maxwellian velocity distribution, conventional astrophysical parameters as well as WIMP coupling to neutrons equal to coupling to protons. Figure~\ref{fig:exclusion} shows the 90\% confidence level projected sensitivity of TREX-DM assuming an energy threshold of 0.4~keVee and a total exposure of 1~kg$\cdot$y in argon and neon, under both a conservative and a realistic value of a flat-shaped background level, 100~keV$^{-1}$kg$^{-1}$d$^{-1}$ and 1~keV$^{-1}$kg$^{-1}$d$^{-1}$, respectively. We have also computed the sensitivity under a more optimistic scenario in which we assume a threshold of 100~eV and a background level of 0.1~keV$^{-1}$kg$^{-1}$d$^{-1}$. If these hypotheses are realized, TREX-DM would have a very competitive sensitivity to low-mass WIMPs, in particular improving the current best limits in the range of $2<m_{\chi}< 8$~GeV, and largely surpassing the region of parameter space ``hinted'' by the positive interpretations of the few dark matter experiments mentioned in a previous section. A word of caution about these sensitivity projections is in order: given that flat backgrounds are assumed, they are critically determined by the background level in the lowest energy bin above threshold. Experimental validation of background levels at such low energies is a must to add fidelity to those projections.


\section{Discussion and outlook}
\label{sec:discussion}

The exploration of ultra-low backgrounds below $\sim$10~keV, and especially at sub-keV energies, is a relatively little-trodden path. Only recent technologies like low threshold (e.g. Luke-Naganov) bolometers, semiconductor detectors with special low-capacitance electrodes, or large arrays of small semiconductor cells (i.e. arrays of low-noise CCDs), show promise of achieving an energy threshold well below 1~keV to events happening in a relatively large sensitive mass. These techniques are being explored mainly in the context of low-mass WIMPs searches, following the motivations discussed in section~\ref{sec:wimps}. The target masses achieved by these novel detectors are still small (below $\sim$1~keV) compared to mainstream WIMP detectors. Also, the normalized background levels achieved are still modest (around a few counts~keV$^{-1}$kg$^{-1}$d$^{-1}$) when compared with the extraordinarily low levels that electron-nuclear discrimination capability provides in leading mainstream WIMP experiments. The prospects of the new low-energy threshold techniques to continuously decrease their background levels will rely on the success of making those techniques sufficiently radiopure (identify materials and components that contain radioactive elements, and find radiopure substitutes for them). In some sense, the new low-threshold--low-background scenario is similar to that of the early days of WIMP searches, in which scrupulous detector radiopurity was the essential, if not the only, handle to get low background. In this case there is the additional difficulty of determining the background sources and phenomenology that affects the low energy (keV and below) window of the detectors. The experimental exploration of the detector background in this regime is \textit{per se} a highly interesting quest.

The results presented in the previous sections of this and the companion paper support our claim that gaseous TPCs read by microbulk Micromegas readouts are a detection technique of high potential for low-background operation at keV energies. Gaseous detectors are naturally suited for low threshold application thanks to the intrinsic amplification happening in gas. Indeed, without the need to resort to gas mixtures optimized for high amplification, highly-pixelised Micromegas planes read out with state-of-the art TPC DAQ electronics, can record signals below the keV, and probably down to 100~eV (see Figure~\ref{fig:threshold}). Although experimental demonstration of this latter fact in the fully equipped TREX-DM prototype is still pending (to be done in the near future once the new DAQ is implemented, see section~\ref{sec:trexdm}), preliminary evidence for this has already been presented~\cite{garza_mpgd_trieste,Iguaz:2015myh}. TPCs in particular enjoy a number of additional features of interest, not available in competing techniques. Signal generation in TPCs happens locally at the readout plane. This means that \textit{a priori}, one can have arbitrarily long drift distances (and therefore arbitrarily large detectors) with no effect in the signal quality (e.g. signal-to-noise ratio) at the readout\footnote{The signal quality will depend on the geometrical features of the readout, practically the pixel/strip capacitance, and the amplification gap. Provided the number of channels scale appropriately with the detector size, keeping the same capacitance per channel, the energy threshold will remain low. With the granularity capabilities of Micromegas and modern high channel density TPC electronics, this does not pose a technical limitation to scaling-up.}. This is not the case for detectors in which signal generation happens in the bulk volume of the detector, for which signal-to-noise ratio is inversely dependent on the detector size, i.e. larger detectors usually have higher threshold. Modern MPGD like Micromegas are now becoming a mature technique that allow to overcome many of the traditional drawbacks of conventional TPCs. Large area readouts are feasible, simple to fabricate and cost-effective. They can be patterned with high granularity, keeping a low capacitance per signal channel. Although topological discrimination at very low energies is expected to be only of mild effect (as low energy events --both signal and background-- are necessarily feature-less), one should not underestimate the topological information as a handle to understand the origin and nature of potentially unexpected background sources that may appear (e.g. special kind of electronic noise) once sufficiently long exposure is reached. Finally, and most importantly, Micromegas readouts of the microbulk type can be fabricated with very high radiopurity. As described in section 3 of the companion paper, bounds to their U and Th intrinsic contamination have been brought down to $\sim$0.1~$\mu$Bq/cm$^2$ levels. Realistic scaling-up strategies, like the ones defined in section 7 of~\cite{trexbbreview}, allow for large area readout implementation with radiopure extraction of high number of signals, altogether avoiding soldering, connectors or any other material or component apart from the microbulk foil itself (and its pure-copper support mechanics). This simplicity in its implementation is a strong point that facilitates future additional background reduction and detector scaling-up.

In the previous sections we have discussed the potential impact in the fields of solar axion detection with axion helioscopes, as well as the search for low-mass WIMPs underground. Although sufficient evidence has been obtained supporting this potential, important steps need to be done in the near future to validate some of the results. On one hand, as part of the IAXO Technical Design Report, the IAXO-D0 setup is being built to demonstrate the current hypothesis to explain the background limitations in current CAST detectors and reach the required background levels for the new project. On the other hand, the TREX-DM prototype described in section~\ref{sec:trexdm} will soon be commissioned underground to prove experimentally the levels anticipated by simulations in section~\ref{sec:prospects}. TREX-DM could presumably reach an exposure at the level of $\sim$1~kg-year. If background levels as anticipated in section~\ref{sec:prospects} are realized, competitive sensitivity to low-mass WIMPs as illustrated in Figure~\ref{fig:exclusion} are already possible with such detector. If even lower background levels are achieved, something a priori perfectly possible, by virtue of the arguments exposed in section~\ref{sec:prospects}, larger exposures (and thus a larger detector) would seem desirable, to reach even better sensitivity. The good scaling-up prospects of this detector configuration exposed above make this a feasible possibility in the mid-term.

Large TPCs with low background in the few keV regime are also a requirement in a new type of axion helioscopes recently studied~\cite{Galan:2015msa}. In this detector configuration, the gas of the TPC plays the role of buffer gas to tune axion-to-photon conversion for particularly high axion masses. The full TPC is placed inside a strong magnetic field, and the solar axions convert and get subsequently absorbed in it. Contrary to conventional axion helioscopes like CAST or IAXO, the detection process is not coherent, and the conversion probability is just proportional to the magnetic volume available. Because of this, there is no privileged direction in the experiment, and there is no need to track the Sun (no need to point/move the magnet) and thus a geometry of a large magnetic TPC is feasible. The lack of coherence is compensated by going to relatively large detector volume. The signal appears as a large O(1) modulation due to the dependency of the conversion on the angle between magnetic field and axion incoming direction. The concept has been shown to be competitive with conventional helioscopes for relatively high axion masses, from 0.1~eV to 10~eV~\cite{Galan:2015msa}. The work here presented may be instrumental to realize this concept in the future.

\section{Conclusions}
\label{sec:conclusion}

We have studied the capabilities of microbulk Micromegas TPCs to achieve very low background levels in the few keV energy regime, with the aim of applying them in the search for solar axions (as part of an axion helioscope setup) and the search for low-mass WIMPs. In both these fields, very promising prospects appear to be open. In the first case, this type of detectors have been used and developed in the context of the CAST experiment at CERN. The last generation of CAST detectors here reviewed have achieved record background levels below $10^{-6}$~\ckcs. Coupled with an x-ray focusing optic, this level corresponds to an effective background of $\sim$1 count in about four months of experimental running time. Evidence that this background is still limited by cosmic-related events has been presented, in particular by the fact that a similar detector system running in an underground location at the LSC has achieved an even lower level of $10^{-7}$~\ckcs. There are prospects to reach this level at surface, and even to improve them further (maybe down to $10^{-8}$~\ckcs). This is the goal of the IAXO-D0 demonstrator under construction as part of the Technical Design Report of the IAXO experiment, a forth generation axion helioscope and follow-up of CAST, currently under preparation.

The low background capabilities of these detectors at the keV scale could also have an impact in the search for low-mass WIMPs, potentially composing our galactic dark matter halo. The TREX-DM prototype to test the concept has been presented. In some sense, TREX-DM is a 10$^3$ times scaled-up version of the CAST/IAXO Micromegas TPCs. The detector is built with state-of-the-art radiopurity specifications, can operate with different gases (most relevantly Ar or Ne mixtures) at high pressures (up to 10~bar), and with energy threshold well below $\sim1$~keV, possibly down to 100~eV. The anticipated background levels from the known intrinsic radioactivity of the detector point to extremely low levels down to, or even below, 1 count~keV$^{-1}$~kg$^{-1}$~d$^{-1}$. The measurement of the detector background at these low energies with this sensitivity in an underground location is highly motivated, as very few such measurements exist. If these values are experimentally confirmed in a planned forthcoming data taking campaign underground, TREX-DM could have very competitive sensitivity to low-mass WIMPs, in particular going beyond current experimental bounds in the 2--8 GeV mass range.

\acknowledgments

The authors would like to warmly acknowledge the many collaborators that have contributed to the T-REX results obtained over the last years, from the GIFNA group, from CEA/Saclay, CERN and LSC, as well as form the CAST, IAXO, NEXT, RD-51 and PandaX-III collaborations for many motivating discussions. Our special thanks go also to Rui de Oliveira workshop at CERN for fabricating for us the microbulk readouts, and CEA/Saclay colleagues for their always wise advice. We acknowledge the use of the Servicio General de Apoyo a la Investigación-SAI, Universidad de Zaragoza. All this research has been mainly supported by the ERC Starting Grant T-REX ref. ERC-2009-StG-240054 of the IDEAS program of the 7th EU Framework Program. Support is acknowledged also from the Spanish Ministry of Economy and Competitiveness under grants FPA2008-03456, FPA2011-24058, and FPA2013-41085-P, and the University of Zaragoza under grant JIUZ-2014-CIE-02. F.I. acknowledges the support from the \emph{Juan de la Cierva} program and T.D from the \emph{Ram\'on y Cajal} program of the MICINN.

\bibliography{trexdm}
\bibliographystyle{JHEP}

\end{document}